\documentclass[journal]{IEEEtran}
\usepackage{amsmath,amsfonts}
\usepackage{amsfonts,amssymb}
\usepackage{algorithmic}
\usepackage{mathrsfs}
\usepackage[ruled,linesnumbered]{algorithm2e}
\usepackage{array}
\usepackage{xcolor}
\usepackage{textcomp}
\usepackage{stfloats}
\usepackage{url}
\usepackage{verbatim}
\usepackage{graphicx}
\usepackage{cite}
\usepackage{bm}
\usepackage{float}
\usepackage{subfigure}
\newtheorem{theorem}{Theorem}

\newenvironment{remark}         {\begin{sideremark}\rm}{\end{sideremark}}
\newtheorem{sideremark}{Remark}
\newcommand{\enc}[1]{\ensuremath{\mathsf{Enc}\left(#1\right)}}
\newcommand{\dec}[1]{\ensuremath{\mathsf{Dec}\left(#1\right)}}
\begin{document}

\title{A Privacy-Preserving Framework for Cloud-Based HVAC Control}
 \author{Zhenan Feng and Ehsan Nekouei 
\thanks{Zhenan Feng and Ehsan Nekouei are with the Department of Electrical Engineering, City University of Hong Kong. E-mail: {\tt zhenafeng2-c@my.cityu.edu.hk, enekouei@cityu.edu.hk}. The work was supported by the Research Grants Council of Hong Kong under Project CityU 21208921 and a grant from Chow Sang Sang Group Research Fund sponsored by Chow Sang Sang Holdings International Limited. }}


\markboth{Journal of \LaTeX\ Class Files,~Vol.~14, No.~8, August~2021}%
{Shell \MakeLowercase{\textit{et al.}}: A Sample Article Using IEEEtran.cls for IEEE Journals}


\maketitle

\begin{abstract}
The objective of this work is $(i)$ to develop an encrypted cloud-based HVAC control framework to ensure the privacy of occupancy information, $(ii)$ to reduce the communication and computation costs of encrypted HVAC control. Occupancy of a building is sensitive and private information that can be accurately inferred by cloud-based HVAC controllers using HVAC sensor measurements, ($iii$) to reduce the leakage of private information via the triggering time instances in event-based encrypted HVAC control systems. To ensure the privacy of the privacy information, in our framework, the sensor measurements of an HVAC system are encrypted by a fully homomorphic encryption technique prior to communication with the cloud controller. We first develop an encrypted fast gradient algorithm that allows the cloud controller to regulate the indoor temperature and CO$_2$ of a building by solving two model predictive control problems using encrypted HVAC sensor measurements. We next develop an event-triggered control policy to reduce the communication and computation costs of the encrypted HVAC control. We cast the optimal design of the event-triggering policy as an optimal control problem wherein the objective is to minimize a linear combination of the control and communication costs. Using Bellman's optimality principle, we study the structural properties of the optimal event-triggering policy and show that the optimal triggering policy is a function of the current state, the last communicated state with the cloud, and the time since the last communication with the cloud. We also show that the optimal design of the event-triggering policy can be transformed into a Markov decision process by introducing two new states. As the triggering time instances are not encrypted, there is a risk that the cloud may use them to deduce sensitive information. To mitigate this risk, we introduce two randomized triggering strategies that reduce the leakage of private information via the triggering time instances. We finally study the performance of the developed encrypted HVAC control framework using the TRNSYS simulator. Our numerical results show that the proposed framework not only ensures efficient control of the indoor temperature and CO$_2$ but also reduces the computation and communication costs of encrypted HVAC control by at least 60\%. 
\end{abstract}


\section{Introduction}
\subsection{Motivation}
Cloud-based control architecture has been widely used in different applications, such as building automation \cite{goldschmidt2015cloud,hitachi,DRGONA202063,TAHERI2024118270}. In this approach, the controller resides in a cloud computing unit and receives the sensor measurements from the system. It then computes the control input using the sensor measurements and sends the control input to the system. Cloud-based control systems have desirable properties such as fast scalability, high computational performance, and high degrees of flexibility and accessibility \cite{r1}. 

The key motivation for adopting a cloud-based control architecture is its suitability and commercial appeal for the control-as-a-service model \cite{DRGONA202063}. Cloud platforms offer on-demand access to computational and storage resources, enabling instant provision of necessary resources for tasks such as computationally intensive HVAC control algorithms. This approach is more efficient and cost-effective than updating the local computing system, making it easier to replicate and deploy different control strategies across large building portfolios.
 However, privacy is a significant concern for the users of cloud-based control systems, as cloud controllers may act as honest but curious adversaries and attempt to infer private information based on the sensor measurements of a system.

Occupancy of a building is a sensitive source of information in building automation applications as it indicates whether a building is vacant or not. Also, the occupancy can be used to track the location of individuals in a building \cite{jiang2016indoor}. However, the occupancy of a building can be accurately inferred from its temperature and CO$_2$ measurements \cite{occlocation}. Hence, a cloud-based HVAC control system can potentially infer the occupancy or track the location of individuals inside a building, which indicates the violation of the occupants' privacy. 

\begin{figure}[t]
    \centering
    \includegraphics[width=3in]{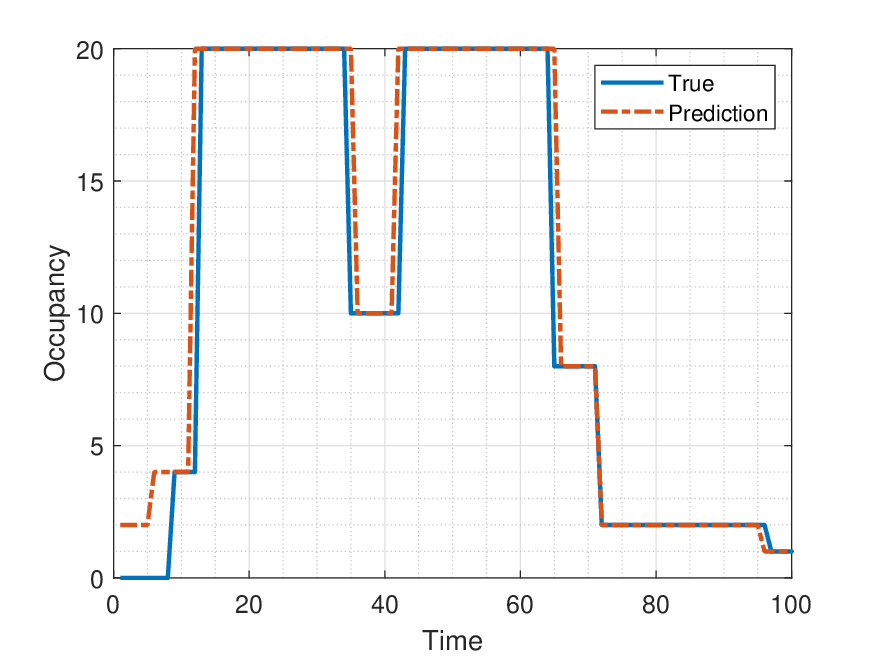}
    \caption{The comparison of true occupancy and predicted building occupancy over time.}
    \label{occupancy}
\end{figure}

 To demonstrate the potential privacy violation in a cloud-based HVAC control system, we performed an experiment wherein a neural network-based occupancy estimator was trained for a building using its CO$_2$ measurements and occupancy data. In our experiment, the CO$_2$ measurements were generated using the TRNSYS simulator. Fig. \ref{occupancy} shows trajectories of the true occupancy and the predicted occupancy by the \textcolor{black}{neural network} occupancy estimator. Based on this figure, the true occupancy of the building can be closely tracked using the CO$_2$ measurements. According to our experiment, the occupancy can be estimated with an accuracy of $92.1\%$ using the sensor measurements.

Homomorphic encryption (HE) is a potential solution for ensuring privacy in cloud-based control applications, as it allows a cloud controller to perform computations using encrypted sensor measurements. In this approach, the sensor measurements of the system are first encrypted using a HE technique, and the cloud controller receives the encrypted sensor measurements. Thus, the cloud controller cannot infer sensitive information from the encrypted measurements.

Although HE is capable of ensuring privacy in cloud-based control, it suffers from a number of drawbacks. First, only addition and multiplication operations can be performed over the data encrypted using HE. Thus, one must first develop an encrypted control law based on the addition and multiplication operations. Second, computations over encrypted data require more computational resources than computations over plaintext data. This would result in a high computational cost when computing-as-a-service \cite{ibm} platforms are utilized for cloud-based control, as the cost of using these platforms depends on the utilized computational resources. Third, encryption increases the data rate between the system and the cloud. This is particularly challenging when wireless sensors are used in the system, since the communication unit of a sensor accounts for a significant part of sensor power consumption. Thus, encrypted communication will significantly reduce the lifetime of a sensor's battery.  

\subsection{Contributions}
In this paper, we developed an encrypted cloud-based HVAC control system using the HE technique to ensure the privacy of the building's occupants. We also develop an optimal event-triggered control policy to reduce the computation and communication requirements of the HE. Our main contributions are summarized as follows:

\begin{enumerate}
    \item We develop a privacy-preserving cloud-based HVAC control framework. To ensure the occupants' privacy, in our framework, the HVAC sensor measurements are first encrypted using a \textcolor{black}{homomorphic} encryption scheme. The cloud receives encrypted sensor measurements and computes the control input (in the encrypted form) for regulating the temperature and CO$_2$ by solving two model predictive control problems. 
    
    \item We design an encrypted fast gradient algorithm that allows the cloud to compute the (encrypted) control input using encrypted sensor measurements. The cloud then transmits the encrypted control input to the HVAC system, which is executed after decryption.  
    
    \item To reduce the communication and computation burdens of the encrypted HVAC control, we develop an optimal event-trigger framework that reduces the frequency of communication between the system and the cloud. We cast the optimal design of the event-triggering unit as an optimal control problem, where the objective is to minimize a linear combination of the control and communication costs. We derive the Bellman optimality principle for the optimal event-trigger design problem and show that the optimal triggering policy only depends on the current state, the last communicated state to the cloud, and the time since the last communication between the system and the cloud.
    
    \item To prevent the cloud from deducing private information using the triggering time instances, \emph{i.e.}, timing information, we introduced two randomized triggering strategies: the $\epsilon$-randomized policy and the entropy-regularized approach for designing the event-triggering unit. Under the randomized triggering policies, it is impossible for an adversary to determine whether the triggering event was made randomly by the event-triggering unit or resulted from a change in the building environment.
    
    \item Using TRNSYS simulator, we study the performance of the developed encrypted event-triggered HVAC control framework in regulating the temperature and CO$_2$ in a building. Our results show that \textcolor{black}{encrypted event-triggered HVAC control can reduce the communication frequency between the system and cloud by more than $60$\%, with a negligible impact on the closed-loop performance, which indicates a significant reduction in the communication and computation costs of encrypted HVAC control.}
\end{enumerate}

\subsection{Related Work}
Homomorphic encryption (HE) refers to an encryption technique that allows computations on encrypted data without using decryption first \cite{rivest1978data}. This technique ensures that the computation using the encrypted data generates a result that is identical to the result obtained by performing the same operation on the unencrypted data \cite{acar2018survey}. HE has been used in secure networked control applications, \emph{e.g.}, see \cite{r2} and references therein. In \cite{r4}, the authors proposed an encrypted linear control using the public key encryption and the ElGamal homomorphic encryption, which is a multiplicative homomorphic encryption \cite{elgamal1985public}. The asymptotic stability of a linear system under a dynamic ElGamal system was studied in \cite{Stability-ElGamal}. The authors in \cite{Kawase-parially} studied the design of dynamic quantizers for encrypted state feedback control of linear systems using the ElGamal encryption scheme.

The authors in \cite{r5} and \cite{r6} studied secure control of linear systems using the additive homomorphic property of the Paillier encryption \cite{paillier1999public}. 
Kim \textit{et al.} \cite{kim2016encrypting} developed a secure control framework for linear systems using a fully homomorphic encryption (FHE) scheme. Zhang \textit{et al.} \cite{Securestate} developed a hybrid encrypted state estimation framework based on two partially HE schemes. We note that the encrypted model-based control has been studied in \cite{lin2018secure,darup2017towards,r8}. The authors in \cite{lin2018secure} designed an encrypted controller for a nonlinear system and studied its closed-loop stability under the proposed controller. 
Darup \textit{et al.} proposed encrypted model predictive control frameworks for linear systems using the Paillier encryption scheme in \cite{darup2017towards} and \cite{r8}.

\begin{figure*}[t]
    \centering
    \includegraphics[width=5.5in]{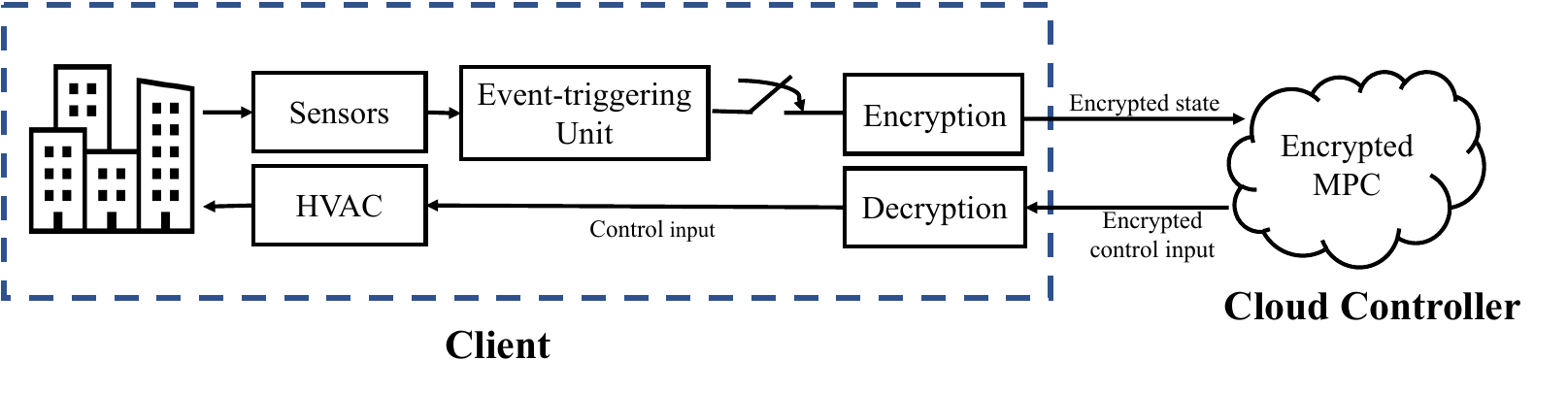}
    \caption{The proposed encrypted cloud-based HVAC control scheme.}
    \label{event-trigger scheme}
\end{figure*}


\subsection{Organization}
Section II introduces the proposed encrypted HVAC control framework. Section III introduces the optimal design of the event-triggering unit. Section IV numerically evaluates the performance of the proposed encrypted event-triggered HVAC control framework using TRNSYS simulator. Finally, Section V concludes the paper.

\section{Encrypted Model Predictive HVAC Control}
In this paper, we develop a privacy-preserving cloud-based HVAC control framework, as shown in Fig. \ref{event-trigger scheme}. In this framework, a cloud-based controller computes the required control inputs to regulate the CO$_2$ and temperature by solving two model predictive control (MPC) problems. To ensure the privacy of the building occupants, the temperature and CO$_2$ measurements are encrypted by a FHE scheme prior to communication with the cloud. We design an encrypted fast gradient algorithm that allows the cloud to compute the control input based on encrypted measurements. To reduce the high communication and computation costs of the FHE, an event-triggering unit is designed to reduce the communication frequency between the HVAC system and the cloud. Thus, the (measurement) communication and (control input) computations will be performed only at specific triggering instances, decided by the event-triggering unit.

In this section, we first describe the building model. We then develop an encrypted optimization algorithm for solving the temperature and CO$_2$ regulation problems using encrypted measurements. In the next section, we study the optimal design of the event-triggering unit.

\subsection{Building Model}
\subsubsection{Thermal Model}We use an RC network to model the thermal dynamics of a building, where walls and rooms are modeled as separate nodes. The dynamic of temperature can be expressed using the following state-space equations 
\begin{align}
    \label{thermal}
    \bm{T}_{t+1} = \bm{A}^{\rm tem}\bm{T}_t + \bm{B}^{\rm tem}    
    \begin{bmatrix}
    \bm{m}_{t} \\\bm{0}
    \end{bmatrix}
    \odot(\bm{T_a}-\bm{T}_t) + \bm{w}_t^{\rm tem},
\end{align}
where $ \bm{T}_t =
    \begin{bmatrix}
    T_r^{i,t}\\T_w^{j,t}
    \end{bmatrix}
    _{i,j}$ is the state of the system, $T_r^{i,t}$ is the temperature of room $i$ at time $t$, and $T_w^{j,t}$ is the temperature of wall $j$ at time $t$. In \eqref{thermal}, $\bm{m}_t$ is the vector of control input whose elements are the air mass flow values into each thermal zone, $\bm{T_a}$ represents the vector of supply air temperature for different zones of the building, $\bm{0}$ is a vector of zeros, size of $\bm{0}$ is equal to the number of walls, and $\odot$ represents the component-wise product of two vectors. The variable $\bm{w}_t^{\rm tem}$ captures the disturbance due to the outside temperature at time $t$, and the matrices $\bm{A}^{\rm tem}$ and $\bm{B}^{\rm tem}$ are the time-invariant parameters of the RC thermal model. We refer the reader to Appendix \ref{App: BM} for a detailed derivation of the thermal model.

The expression for the thermal model described in (\ref{thermal}) is nonlinear due to the product of control input and state variables. To linearize (\ref{thermal}), we use the following change of variables
\begin{equation}
    \label{feed1}
    \bm{u}^{\rm tem}_{t} =    
    \begin{bmatrix}
    \bm{m}_{t} \\\bm{0}
    \end{bmatrix}
    \odot(\bm{T_a}-\bm{T}_t),
\end{equation}
where \textcolor{black}{$\bm{u}^{\rm tem}_{t}$} is the vector of new control variables at time $t$ after feedback linearization. Substituting (\ref{feed1}) in (\ref{thermal}), we obtain
\begin{equation}
    \label{feedthermal}
    \bm{T}_{t+1} = \bm{A}^{\rm tem}\bm{T}_t + \bm{B}^{\rm tem}\bm{u}^{\rm tem}_{t} + \bm{w}_t^{\rm tem}.
\end{equation}

\subsubsection{CO$_2$ Model}The evolution of CO$_2$ in zone $i$ can be expressed as 
\begin{equation}
\label{CO2}
\frac{dC_t^{i}}{dt}=\left(C_{o}-C_t^i\right)\frac{m^i_t}{V}+\frac{G^i_t}{V}\nonumber,
\end{equation}
where $C_t^i$ is the indoor CO$_2$ concentration of room $i$ at time $t$, $C_{o}$ is the outdoor CO$_2$ concentration, $m^i_t$ is the current mass flow rate of room $i$, and $G^i_t$ is the amount of CO$_2$ generated by human occupants at time $t$, and $V$ is room volume. The above differential equation can be transformed into the following state space form
\begin{equation}
    \label{stateCO2}
    \bm{C}_{t+1} = \bm{A}^{\rm c}\bm{C}_t + \bm{B}^{\rm c}\frac{\bm{m}_t}{V}\odot(\bm{C}_{o}-\bm{C}_t) + \bm{w}_t^{\rm c}\nonumber,
\end{equation}
where $\bm{C}_t=\left[C^i_t\right]_i$ is the system state, and $\bm{w}_t^{\rm c}$ is the disturbance due to the CO$_2$ generated by the building occupants.  Using the change of variables  $\bm{u}^{\rm c}_{t} = \frac{\bm{m}_t}{V}\odot(\bm{C}_{o}-\bm{C}_t)$, the  CO$_2$ dynamic can be transformed into the following linear form
\begin{equation}
    \label{feedCO2}
    \bm{C}_{t+1} = \bm{A}^{c}\bm{C}_t + \bm{B}^{c}\bm{u}^{\rm c}_t + \bm{w}_t^{\rm c}.
\end{equation}

Combining \eqref{feedthermal} and \eqref{feedCO2}, the evolution of the temperature and CO$_2$ in the building can be expressed as   \begin{align}
    \bm{x}_{t+1} = \bm{Ax}_t + \bm{Bu}_t + \bm{w}_t,\nonumber
\end{align}
where $\bm{x}_{t} =
    \begin{bmatrix}
    \bm{T}_t\\\bm{C}_t
    \end{bmatrix}$, $\bm{w}_{t} =
    \begin{bmatrix}
    \bm{w}^{\rm tem}_t\\\bm{w}^{\rm c}_t
    \end{bmatrix}$ and $\bm{u}_t=\begin{bmatrix}
    \bm{u}^{\rm tem}_t\\\bm{u}^{\rm c}_t
    \end{bmatrix}$.
\subsection{Encrypted HVAC control}
The privacy of the building occupants becomes a major concern when an untrusted controller, \emph{e.g.}, a cloud-based controller, is used to regulate the temperature and CO$_2$ in different zones.
To ensure the privacy of the building's occupants, we propose an encrypted model predictive control (MPC) approach for controlling the HVAC system. In this framework, the temperature and CO$_2$ measurements are encrypted using the CKKS \cite{cheon2017homomorphic} encryption scheme, and the controller receives the encrypted measurements and computes the required air mass flow for regulating the temperature and CO$_2$ by solving two MPC problems. The controller then transmits the (encrypted) control inputs to the HVAC system which are executed by the actuators after decryption. 


\subsubsection{Encrypted temperature regulation}
We first present the optimization problem for temperature regulation. We then present an encrypted optimization algorithm that allows the controller to solve the temperature regulation problem using encrypted temperature measurements. 

To regulate the temperature, the cloud solves the following MPC problem
\begin{equation}
    \label{MPC}
    \begin{split}
    \min_{{\left\{u^{\rm tem}_{t+k|t}\right\}_{k=0}^{T-1}}}{\frac{1}{T}}\sum_{i=0}^{T-1} & {(\bm{T}_{t+i+1|t}-\bm{T}_{ref})^\mathrm{T}\bm{Q}(\bm{T}_{t+i+1|t}-\bm{T}_{ref}) } \\
     & +{\bm{u}^{\rm tem}_{t+i|t}}^\mathrm{T}\bm{R}{\bm{u}^{\rm tem}_{t+i|t}}, \\
     \text{s.t.} \quad \bm{T}_{k+1|t} & =\bm{A}^{\rm tem}\bm{T}_{k|t}+\bm{B}^{\rm tem}\bm{u}^{\rm tem}_{k|t}+\bm{M}_k^{\rm tem},\ \ k\geq t \\
    \bm{u}^{tem}_{i|t} & \in\mathcal{U}^{tem},\ \ i=t,t+1,\ldots,t+T-1 \\
    \bm{T}_{t|t} & =\bm{T}_t,
    \end{split}
\end{equation}
where $\bm{T}_t$ is the actual temperature of different zones at time $t$, $\bm{T}_{k+1|t}$ is the predicted temperature based on the RC model, $\bm{u}^{\rm tem}_t$ is the control input of the linearized temperature model, $T$ is the horizon length, $\bm{M}_k^{tem}$ is the mean of daily ambient temperature at time $k$, $\bm{Q}$ and $\bm{R}$ are positive definite weight matrices, and $\bm{T}_{ref}$ is the vector of temperature reference, which has the same dimension as $\bm{T}_t$. In \eqref{MPC}, the constraint set $\mathcal{U}^{tem}$ is in the form of $\mathcal{U}^{tem}=\left\{l_m\leq u^{tem}_t\leq h_m\right\}$ where $l_m$ and $h_m$ are vectors of appropriate dimensions.

To solve the MPC problem in (\ref{MPC}), we transform it into the following equivalent optimization problem
\begin{equation}
    \label{quadratic}
   \mathop{\min}\limits_{\bm{u}_t^{\rm tem}}{\frac{1}{T}}\left({\bm{u}_t^{\rm tem}}^\mathrm{T}\bm{H}\bm{u}_t^{\rm tem}+2{\bm{u}_t^{\rm tem}}^\mathrm{T}\bm{F}^\mathrm{T}{\bm{u}_t^{\rm tem}}\bm{T}_t\right),
\end{equation}
where $\bm{u}_t^{\rm tem}=\left[\bm{u}^{\rm tem}_{t|t},\bm{u}^{\rm tem}_{t+1|t},\ldots,\bm{u}^{\rm tem}_{t+T-1|t}\right]^\mathrm{T}$ and matrices $\bm{H}$ and $\bm{F}$ can be obtained using the method described in \cite{borrelli2017predictive}. The optimization problem \eqref{quadratic} can be solved with the projected fast gradient method (FGM) \cite{nesterov2003introductory} in Algorithm \ref{FGM-Plain} where $\bm{T}_t$ is the temperature of different zones at time $t$, $u_t^{\rm tem}(k)$ is the value of the optimization variable ($u_t^{\rm tem}$) at iteration $k$, $u_t^{\rm tem}(0)$ is the initial value of $u_t^{\rm tem}(k)$, $\lambda_{max}(\bm{H})$ is the largest eigenvalue of $\bm{H}$, ${\kappa}(\bm{H})$ is the condition number of $\bm{H}$ \cite{borrelli2017predictive,kempf2020fast}, $h^i_m$ is the $i$th element of $h_m$, $l^i_m$ is the $i$th component of $l_m$, and $u_{t}^{{\rm tem},i}(k+1)$ is the $i$th element of $u_t^{\rm tem}(k)$. 

\begin{algorithm}[ht]
\caption{Fast gradient method for solving  \eqref{MPC}}\label{FGM-Plain}
\KwIn{$\bm{T}_t,u_t^{\rm tem}(0)$}
\KwOut{$\bm{u}_t^{\rm tem}$}
Set $\xi(0)=u_t^{\rm tem}(0)$\; 
Set $L=\lambda_{max}(\bm{H})$\;
Set  $\eta=(\sqrt{{\kappa}(\bm{H})}-1)/(\sqrt{{\kappa}(\bm{H})}+1)$\;
\For{$k=0$ \KwTo $K-1$}{
  ${d_{t}(k)=(\bm{I}_{n}-\frac{1}{L}\bm{H})\xi_{t}(k)+(-\frac{1}{L} \bm{F}^\mathrm{T} )\bm{T}_t}$\;
  \uIf{$d_{t}^i(k)<l_m^i$}{
		$u_{t}^{{\rm tem},i}(k+1)=l_m^i$\;
	}
  \uElseIf{$d_{t}^i(k)>h_m^i$}{
        $u_{t}^{{\rm tem},i}(k+1)=h_m^i$\;
    }
  \Else{
        $u_{t}^{{\rm tem},i}(k+1)=d_{t+k}^i$\;
    }
  $\xi_{t}(k+1)=(1+\eta)\bm{u}_{t}^{\rm tem}(k+1)+(-\eta)\bm{u}_{t}^{\rm tem}(k)$\;
}
$\bm{u}_t^{\rm tem}=\bm{u}_t^{\rm tem}(K)$
\end{algorithm}

To solve the MPC problem in \eqref{quadratic} using the cloud, first, the measurements of temperature sensors, \emph{i.e.}, $\bm{T}_t$, are encrypted using the CKKS method, and the encrypted sensor measurements are communicated with the cloud. Then, the cloud uses an encrypted version of the Algorithm \ref{FGM-Plain} to solve \eqref{quadratic}. To present the encrypted version of Algorithm \ref{FGM-Plain}, we use the $\enc{}$ and $\dec{}$ to denote the encryption and decryption operations, respectively, by the CKKS scheme. We also use $\otimes$ and $\oplus$ to denote the multiplication and addiction operations of the ciphertext data under the CKKS scheme. The reader is referred to Appendix \ref{App: CKKS} for a brief overview of the encryption, decryption, multiplication, and addition operations under the CKKS encryption scheme. 

Steps $5$ and $13$ in Algorithm \ref{FGM-Plain} only involve addition and multiplication. Thus, they can be performed in the encrypted form by the cloud as follows,
\begin{align}\label{Eq: O1}
 \enc{d_{t}(k)}=&\enc{\bm{I}_{n}-\frac{\bm{H}}{L}}\otimes\enc{\xi_{t}(k)}\nonumber\\ &\oplus\enc{-\frac{\bm{F}^\mathrm{T}}{L}}\otimes\enc{\bm{T}_t},
\end{align}
\begin{align}\label{Eq: O2}
     \enc{\xi_{t}(k+1)}=&\enc{1+\eta}\otimes\enc{\bm{u}_{t}^{tem}(k+1)}\nonumber\\
     &\oplus\enc{-\eta}\otimes\enc{\bm{u}_{t}^{tem}(k)}.
\end{align}
The projection operation in steps $6$-$12$ of Algorithm \ref{FGM-Plain} is a nonlinear operation and cannot be performed using the encrypted data. To perform the projection, the cloud sends the encrypted version of $d_{t}(k)$, \emph{i.e.}, $\enc{d_{t}(k)}$, to the system. The system first decrypts $\enc{d_{t}(k)}$ and computes $u_{t}^{\rm tem}(k+1)$ by performing the projection as follows,
\begin{equation}\label{Eq: EP}
u_{t}^{{\rm tem},i}(k+1)= \left \{
    \begin{array}{ll}
    l_m^i,                    & {d_{t}^i(k)}<l_m^i\\{}
    d_{t+k}^i,     & {d_{t}^i(k)}\in[l_m^i,h_m^i].\\{}
    h_m^i,           & {d_{t}^i(k)}>h_m^i {}
    \end{array}
    \right.
\end{equation}
The system then encrypts $u_{t}^{{\rm tem}}(k+1)$ and sends $\enc{u_{t}^{{\rm tem}}(k+1)}$ to the cloud, which is used to compute the $\enc{\xi_{t}(k+1)}$ using \eqref{Eq: O2}. The encrypted FGM is summarized in Algorithm \ref{FGM-Enc}.

We note that the encrypted FGM in Algorithm \ref{FGM-Enc} requires communication between the system and the cloud due to the nonlinearity of the projection operation. However, the FGM converges fast since optimization problem \eqref{MPC} is quadratic. Thus, the optimal solution of \eqref{MPC} can be found using a few iterations, \emph{i.e.}, a few rounds of communication between the system and the cloud. In practice, an acceptable control performance can be attained by using only one iteration of the encrypted FGM, see subsection \ref{iterationexp}.
\begin{remark}
   The computational cost of encryption can be reduced using a partially homomorphic encryption scheme such as Paillier encryption. However, the cloud will have access to unencrypted system parameters in this case which might be undesirable. 
\end{remark}

\begin{algorithm}[ht]
\caption{Encrypted FGM for solving  \eqref{MPC}}\label{FGM-Enc}
\KwIn{$\enc{\bm{T}_t},\enc{u_t^{\rm tem}(0)}$}
\KwOut{$\enc{\bm{u}_t^{\rm tem}}$}
Set $\xi(0)=\enc{u_t^{\rm tem}(0)}$\; 
Set $L=\lambda_{max}(\bm{H})$\;
Set  $\eta=(\sqrt{{\kappa}(\bm{H})}-1)/(\sqrt{{\kappa}(\bm{H})}+1)$\;
\For{$k=0$ \KwTo $K-1$}{
  Cloud computes $\enc{d_{t}(k)}$ using \eqref{Eq: O1}\;
   Cloud sends $\enc{d_{t}(k)}$ to the system\;
   System decrypts $\enc{d_{t}(k)}$\;
   System computes $u_{t}^{\rm tem}(k+1)$ using \eqref{Eq: EP}\;
   System sends $\enc{u_{t}^{\rm tem}(k+1)}$ to the cloud\;
   Cloud uses $\enc{u_{t}^{\rm tem}(k+1)}$ to compute $\enc{\xi_{t}(k+1)}$ according to \eqref{Eq: O2}.
}
$\enc{\bm{u}_t^{\rm tem}}=\enc{\bm{u}_t^{\rm tem}(K)}$
\end{algorithm}
\subsubsection{Encrypted CO$_2$ regulation}
The CO$_2$ level of each zone is regulated by solving the following optimization problem
\begin{equation}
    \label{MPCCO}
    \begin{split}
    \min_{{\left\{u^{\rm c}_{t+k|t}\right\}_{k=0}^{T-1}}}{\frac{1}{T}}\sum_{i=0}^{T-1} & {(\bm{C}_{t+i+1|t}-\bm{C}_{ref})^\mathrm{T}\bm{Q}(\bm{C}_{t+i+1|t}-\bm{C}_{ref}) } \\
     & +{\bm{u}^{\rm c}_{t+i|t}}^\mathrm{T}\bm{R}\bm{u}^{\rm c}_{t+i|t}, \\
     \text{s.t.} \quad \bm{C}_{k+1|t} & =\bm{A}^{\rm c}\bm{C}_{k|t}+\bm{B}^{\rm c}\bm{u}^{\rm c}_{k|t}+\bm{M}_k^{\rm c},\ \ k\geq t \\
    \bm{C}_{t|t} & =\bm{C}_t, \\
    \bm{u}_{i|t}^{\rm c} & \in\textcolor{black}{\mathcal{U}^{\rm c}},\ \ i=t,t+1,\ldots,t+T-1
    \end{split}
\end{equation}
where $\bm{C}_{k+1|t}$ is the predicted CO$_2$ value, $\bm{C}_t$ is the vector of CO$_2$ concentration in different zones at time $t$, $T$ is the horizon length, $\bm{M}_k^{c}$ is the mean of CO$_2$ generation of occupants, \textcolor{black}{$\mathcal{U}^{\rm c}$ is the constraint set}, $\bm{Q}$ and $\bm{R}$ are positive definite weight matrices. $\bm{C}_{ref}$ is the vector of CO$_2$ concentration reference, which has the same dimension as $\bm{C}_t$.

Similar to the encrypted temperature regulation problem, the CO$_2$ measurements are encrypted using the CKKS methods. The cloud controller computes the required air mass flow for CO$_2$ regulation using the encrypted CO$_2$ measurements and the FGM algorithm. Since we use two separate MPC problems to regulate the temperature and CO$_2$, the HVAC system receives two air mass flow values for each zone from the cloud, \emph{i.e.}, one air mass flow value for the CO$_2$ regulation and one air mass flow for temperature regulation. For each zone, the HVAC system executes the largest air mass flow value at each time step.

\section{Optimal Design of Event-triggering Unit}
To reduce the high communication and computation costs of FHE, we introduce an event-triggering unit to decide whether the system communicates the encrypted measurements with the cloud, as shown in Fig. \ref{event-trigger scheme}. In this section, we first formulate the optimal design of the event-triggering unit as an optimal control problem. We then study the structural properties of the optimal triggering policy. 
 We finally translate the design of the event-triggering unit into a Markov decision process (MDP) and derive Bellman's optimality principle for the optimal triggering policy.

Let $a_t(\mathfrak{I}_t)$ denote the decision of the event-triggering unit at time $t$ where $\mathfrak{I}_t$ is the set of all the information available to the event-triggering unit at time $t$, \emph{i.e.}, $\mathfrak{I}_t=\left\{\bm{x}_0,a_0,\dots,\bm{x}_{t-1},a_{t-1},\bm{x}_t\right\}$. The decision of the event-triggering unit can take two values $a_t(\mathfrak{I}_t)\in\left\{0,1\right\}$. The event-triggering unit transmits the state to the controller when $a_t(\mathfrak{I}_t)=1$. Otherwise, the system states will not be transmitted to the controller. We assume $a_0(\mathfrak{I}_0)=1$ at time $t=0$. 

\begin{figure}[H]
    \centering
    \includegraphics[width=3in]{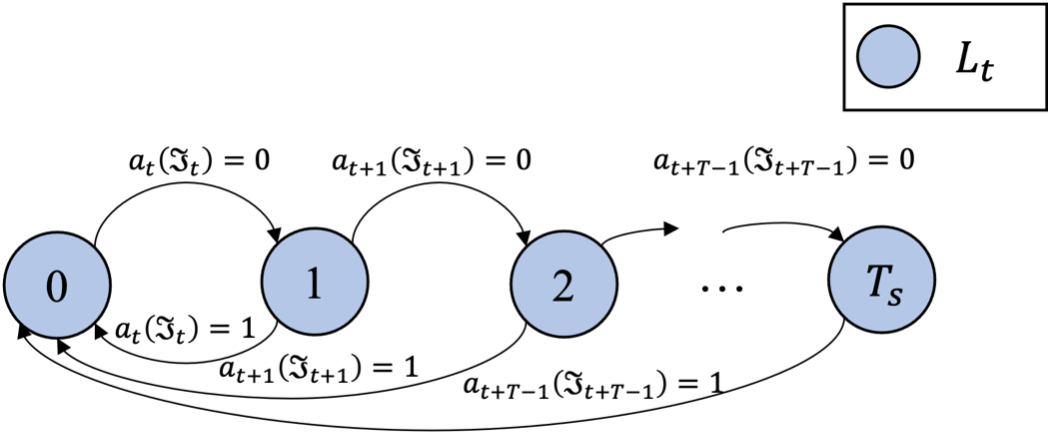}
    \caption{The time evolution of $L_t$.}
    \label{Evolution}
\end{figure}

To prevent the event-triggering unit from always opting out of communication, we impose an upper bound on the time interval between two consecutive triggering instances.  To this end, let $L_t$ denote the number of samples since the last communication between the system and the controller, \emph{i.e.},
\begin{equation}
\label{L_t}
L_t = t - s_t^\star,\nonumber
\end{equation}
where $s_t^\star$ is the last time instance when the event-triggering unit communicated a state to the controller. If $L_t$ is more than the threshold level $T_s$, the event-triggering unit communicates the state to the controller. Fig. \ref{Evolution} shows the time evolution of $L_t$.

 The optimal design of the event-triggering policy can be formulated as the optimal control problem in \eqref{objective function} 
\begin{figure*}
    \begin{align}
    \label{objective function}
     \min_{a_1,...,a_H} &\sum_{t=1}^H E\left[ (\bm{x}_t-\bm{x}_r)^\mathrm{T} \bm{Q} (\bm{x}_t-\bm{x}_r) + \bm{u}_{t}^\mathrm{T} \bm{R} \bm{u}_{t}+ \lambda a_t\right]+ E\left[(\bm{x}_{H+1}-\bm{x}_r)^\mathrm{T} \bm{Q} (\bm{x}_{H+1}-\bm{x}_r)\right].\nonumber  \\
    &\bm{x}_{t+1} = \bm{Ax}_t + \bm{Bu}_t + \bm{w}_t.
    \end{align} 
\hrule
\end{figure*}
where $\lambda$ is a positive number capturing the penalty associated with communication with the cloud; that is, the event-triggering unit receives a penalty equal to $\lambda$ if $a_t=1$. The control input $\bm{u}_t$ can be written as  
\begin{equation}
\label{system state}
\bm{u}_t = (1 - a_t)\bm{u}_{t|s_t^\star} +  a_t\bm{u}_{t|t}.\nonumber
\end{equation}
Note that if the system communicates with the cloud at time $t$, the HVAC system will receive the control inputs $\bm{u}_{t|t},\bm{u}_{t+1|t},\ldots,\bm{u}_{t+T-1|t}$ from the cloud, which are obtained by solving the MPC problems. Thus, the HVAC system executes $\bm{u}_{t|t}$ at time $t$. However, if the system does not communicate with the cloud at time $t$, it has access to the control inputs computed by the cloud in the last communication instance, \emph{i.e.}, $\bm{u}_{s_t^\star|s_t^\star},\bm{u}_{s_t^\star+1|s_t^\star},\ldots,\bm{u}_{s_t^\star+T-1|s_t^\star}$. Thus, the HVAC system executes $\bm{u}_{t|s_t^\star}$ at time $t$.
\subsection{The Sufficient Information for Decision-making}
The computational complexity of the optimal (stationary) event-triggering policy becomes prohibitive as $t$ becomes large. This is due to the fact that the input to the event-triggering policy in \eqref{objective function} is $\mathfrak{I}_t=\left\{\bm{x}_0,a_0,\dots,\bm{x}_{t-1},a_{t-1},\bm{x}_t\right\}$, \emph{i.e.}, the history of information available to the event-triggering unit. Note that the size of $\mathfrak{I}_t$ increases with time. 

In an optimal control problem, 
the optimal policy may only depend on a subset of available information to the decision-maker rather than all the available information. The required information for optimal decision-making is referred to as ``sufficient information". The next theorem derives sufficient information for the optimal event-triggering policy. 
\begin{theorem}\label{Theo: Suff-Inf}
    Let $a_t^\star$ denote the optimal event-triggering policy at time $t$. Then, $a_t^\star$ depends on $\bm{x}_t,\bm{x}_{s_t^\star}, L_t$ where $\bm{x}_t$ is the current state, $\bm{x}_{s_t^\star}$ is the last communicated state with the controller, $L_t$ is the time since last communication between the event-triggering unit and the controller.
\end{theorem}
\begin{IEEEproof}
See Appendix \ref{App: Suff-Inf}.
\end{IEEEproof}
According to Theorem \ref{Theo: Suff-Inf}, $\left\{\bm{x}_t,\bm{x}_{s_t^\star}, L_t\right\}$ is the necessary information for decision-making. This result allows us to restrict the reach for the optimal event-triggering policy to the policies that only depend on $\left\{\bm{x}_t,\bm{x}_{s_t^\star}, L_t\right\}$, rather than $\mathfrak{I}_t$, which significantly simplifies the search for the optimal policy.
\subsection{Dynamic Programming Decomposition}

The next theorem derives the dynamic programming decomposition for the optimal control problem in \eqref{objective function}, which allows us to study the structure of its corresponding optimal value function.

\begin{theorem} \label{Theo: Value_func}
    Let $V_t (\cdot)$ denote the optimal value function at time $t$ associated with the optimal event-triggering policy design problem in \eqref{objective function}. Then, $V_t (\cdot)$ satisfies the following backward optimality equation
    \begin{equation}
V_t (\mathfrak{I}_t^\star) = \min_{a_t}(C_t (\mathfrak{I}_t^\star,a_t ) + E \left\{ V_{t+1} (\mathfrak{I}_{t+1}^\star)|\mathfrak{I}_t^\star,a_t \right\} ), \nonumber
\end{equation}
with the terminal condition 
\begin{equation}
V_{H+1} (\mathfrak{I}^\star_{H+1}) = (\bm{x}_{H+1}-\bm{x}_r)^\mathrm{T} \bm{Q} (\bm{x}_{H+1}-\bm{x}_r),\nonumber
\end{equation}
    where $\mathfrak{I}_{t+1}^\star=\left\{\bm{x}_t,\bm{x}_{s_t^\star}, L_t\right\}$ is the sufficient information for decision-making at time $t$, and 
    \begin{align}
        C_t (\mathfrak{I}_t^\star,a_t )=(\bm{x}_t-\bm{x}_r)^\mathrm{T} \bm{Q} (\bm{x}_t-\bm{x}_r) + \bm{u}_{t}^\mathrm{T} \bm{R} \bm{u}_{t}
 + \lambda a_t \nonumber.
    \end{align}
\end{theorem}
\begin{IEEEproof}
The proof follows from a similar argument as the proof of Theorem \ref{Theo: Suff-Inf}, and is omitted to avoid repetition.
\end{IEEEproof}
Theorem \ref{Theo: Value_func} provides a recursive optimality equation that can be used to compute the optimal value function and the optimal event-triggering policy using value-based methods, such as A2C. Based on this theorem, the optimal value function at time $t$ is not only a function of the state at time $t$, but also the last communicated state with the cloud ($\bm{x}_{s_t^\star}$) as well as $L_t$. Thus, different from the standard MDP formulations where the optimal value function only depends on the current state, in our set-up, any approximation (or representation) of the optimal value function should depend on the current state, the last communicated state with the cloud and $L_t$.

\subsection{Translation into an MDP}
The optimization problem in \eqref{objective function} is not Markovian as the control input at time $t$ will be a function of the state at time ${s_t^\star}$ ($\bm{x}_{s_t^\star}$) rather than $\bm{x}_{t}$ when the event-triggering unit does not communicate with the cloud, \emph{i.e.}, $a_t=0$. This would hinder the application of the existing MDP algorithms for computing the optimal policy. In this section, we show that the optimization problem \eqref{objective function} can be translated into an MDP by introducing two extra states.

Let $\bm{\mathcal{Y}}_t$ denote a new state which is updated according to \begin{equation}
\label{state_y}
\bm{\mathcal{Y}}_{t+1} = (1 - a_t)\bm{\mathcal{Y}}_t + a_t\bm{x}_t,\nonumber
\end{equation}
with the initial condition $\bm{\mathcal{Y}}_1 = \bm{x}_1$. Note that $\bm{\mathcal{Y}}_t$ stores the last communicated state with the cloud. If the event-triggering unit communicates with the cloud at time $t$, we have $a_t=1$, and $\bm{\mathcal{Y}}_{t+1}$ will be \textcolor{black}{equal to $\bm{x}_t$}. However, if the event-triggering unit does not communicate with the cloud at time $t$, we have $a_t=0$ and the value of $\bm{\mathcal{Y}}_{t+1}$ will be equal to the value of $\bm{\mathcal{Y}}_t$. We also use $L_t$ as a new state, which is updated using 
\begin{align}
    L_{t+1}=(1 - a_t)(L_t+1).\nonumber
\end{align}
Thus, when $a_t$ is zero, the value of $L_{t+1}$ is equal to $L_t+1$, and $L_{t+1}$ will be equal to zero if $a_t$ is equal to $1$.

Finally, we augment $\bm{x}_t$ with $\bm{\mathcal{Y}}_t$ and $L_t$ to form a new system whose state updates according to 
\begin{equation}
\begin{aligned}
\begin{bmatrix} \bm{x}_{t+1} \\
\bm{\mathcal{Y}}_{t+1} \\
L_{t+1}\end{bmatrix}= 
\begin{bmatrix} \bm{Ax}_t + \bm{Bu}_{t} + \bm{w}_t \\
(1 - a_t)\bm{\mathcal{Y}}_t + a_t\bm{x}_t \\
(1 - a_t)(L_t+1)\end{bmatrix}.
\end{aligned}
\label{It}
\end{equation}
Note that the dynamics in \eqref{It} is Markovian as $\bm{u}_{t}$ depends on $\bm{x}_{t}$  when $a_t=1$, and depends on $L_t$ and $\bm{\mathcal{Y}}_{t}$ when $a_t=0$.

\subsection{Randomized Triggering Policies}
Since the triggering time instances, \emph{i.e.}, timing information, are not encrypted, the cloud might use them to infer private information. This is because of the potential correlation between the timing information and the change in occupancy. To address this problem, we propose two randomized triggering policies that reduce the risk of privacy loss via timing information: $(i)$ the $\epsilon$-randomized policy, $(ii)$ an entropy-regularized approach.

Under the $\epsilon$-randomized policy, the event-triggering unit with probability $1-\epsilon$ uses the optimal policy $a^{\star}_t$, with probability $\frac{\epsilon}{2}$ decides to send the state to the cloud, and with probability $\frac{\epsilon}{2}$ does not send the state to the cloud.  Let $\pi_\epsilon\left(\cdot\right)$ denote the $\epsilon$-randomized policy. Then, we have  

\begin{equation}\label{greedy}
\pi_\epsilon\left(\mathfrak{I}^{\star}_t\right) =\left \{
\begin{array}{ll}
    a^{\star}_{t}\left(\mathfrak{I}^{\star}_t\right)     & \text{with probability}\;1-\epsilon\\
    1           & \text{with probability}\;\frac{\epsilon}{2}\\
        0           & \text{with probability}\;\frac{\epsilon}{2}.\nonumber
    
\end{array}
\right.
\end{equation}
Under this policy, the adversary does not know whether the triggering decision was randomly taken by the even-triggering unit or it was due to the change in the building environment, \emph{e.g.}, change in occupancy.
    
We also propose an entropy-regularized formulation for the design of the event-triggering unit. Here, we consider the class of stochastic event-triggering policies, in which a policy is represented by the conditional probability distribution $\pi_\theta(a_t|\mathfrak{I}^{\star}_t)$ where $\mathfrak{I}^{\star}_t$ is the necessary information for decision-making in Theorem 1 and $\theta$ is the parameter of the policy. Given $\mathfrak{I}^{\star}_t$, the event-triggering unit randomly decides to transmit with probability $\pi_\theta(a_t =1 |\mathfrak{I}^{\star}_t)$.

The entropy of the distribution $\pi_\theta(\cdot|\mathfrak{I}^{\star}_t)$ is defined as \eqref{entropy}.
\begin{figure*}
    \begin{align}
    \label{entropy}
    J_\theta(\mathfrak{I}^{\star}_t)= -\pi_\theta(a_t=1|\mathfrak{I}^{\star}_t)\log \pi_\theta(a_t=1|\mathfrak{I}^{\star}_t) -\pi_\theta(a_t=0|\mathfrak{I}^{\star}_t)\log \pi_\theta(a_t=0|\mathfrak{I}^{\star}_t).
    \end{align}
    \hrule
    \begin{align}
    \label{new objective function}
    G\left(\theta\right) =&\sum_{t=1}^H E\left[ (\bm{x}_t-\bm{x}_r)^\mathrm{T} \bm{Q} (\bm{x}_t-\bm{x}_r) + \bm{u}_{t}^\mathrm{T} \bm{R} \bm{u}_{t}+ \lambda a_t-\beta J_\theta(\mathfrak{I}^{\star}_t)\right]+ E\left[(\bm{x}_{H+1}-\bm{x}_r)^\mathrm{T} \bm{Q} (\bm{x}_{H+1}-\bm{x}_r)\right].
    \end{align}
    \hrule
    \begin{align}\label{eqgradient}
    &\nabla_\theta     G\left(\theta\right)=\nonumber\\
    & E \left[\left( \sum_{t=1}^{H}(\bm{x}_t-\bm{x}_r)^\mathrm{T} \bm{Q} (\bm{x}_t-\bm{x}_r) + \bm{u}_{t}^\mathrm{T} \bm{R} \bm{u}_{t}+ \lambda a_t-\beta J_\theta(\mathfrak{I}^{\star}_t)+(\bm{x}_{H+1}-\bm{x}_r)^\mathrm{T} \bm{Q} (\bm{x}_{H+1}-\bm{x}_r)\right)\sum_{t=1}^H \nabla_\theta \log\pi_\theta(a_t|\mathfrak{I}^{\star}_t)\right]\nonumber\\
    &+\beta E\left[\sum_{t=1}^H\left(\nabla_\theta\pi_\theta(a_t=1|\mathfrak{I}^{\star}_t)\log \pi_\theta(a_t=1|\mathfrak{I}^{\star}_t) +\nabla_\theta\pi_\theta(a_t=0|\mathfrak{I}^{\star}_t)\log \pi_\theta(a_t=0|\mathfrak{I}^{\star}_t)\right)\right].
    \end{align}
\hrule
\end{figure*}
Discrete entropy is always non-negative and is zero for deterministic policies. Thus, to obtain a randomized event-triggering policy, we add entropy as a regularization term to the objective function of the event-triggering design problem as shown in \eqref{new objective function} where $\beta$ is a positive constant. Then, the optimal entropy-regularized event-triggering policy is the solution of the following optimization problem 
\begin{equation}\label{Eq: Entropy-Penalized}
\min_{\theta} G\left(\theta\right). 
\end{equation}
Similar to the $\epsilon$-randomized method, entropy regularization results in a stochastic event-triggering policy and reduces the leakage of private information via timing data. However, the randomization depends on the state of the system under the entropy-regularized policy, whereas under the $\epsilon$-randomized policy, the randomization is independent of the state.

The optimization problem \eqref{Eq: Entropy-Penalized} cannot be solved using the standard policy gradient theorem \cite{sutton1999policy} since the objective function in \eqref{new objective function} is a nonlinear function of the policy. To tackle this problem, we derive an expression for the gradient of the objective function in \eqref{Eq: Entropy-Penalized} with respect to $\theta$ in the next theorem.
\begin{theorem}\label{gradient}
  The gradient of the objective function in \eqref{Eq: Entropy-Penalized} with respect to $\theta$ can be written as \eqref{eqgradient}.
\end{theorem}
\begin{IEEEproof}
See Appendix \ref{App: gradient}.
\end{IEEEproof}
Theorem \ref{gradient} can be used to find an optimal entropy-penalized policy using the stochastic gradient descent algorithm. According to Theorem \ref{gradient}, the gradient of $G(\theta)$ consists of two terms. The first term follows from the standard policy gradient theorem, and the second term captures the impact of entropy penalization on the gradient. Note that the second term does not appear in the standard policy gradient theorem as the objective function is linear in policy, whereas the objective function in \eqref{Eq: Entropy-Penalized} is nonlinear in the policy $\pi_\theta$ due to the entropy term. Thus, the application of standard policy gradient algorithm to optimization problem \eqref{Eq: Entropy-Penalized} will not result in the correct solution.

\section{Numerical Results}
\subsection{Simulation Set-up}
\begin{figure}[H]
    \centering
    \includegraphics[width=2.6in]{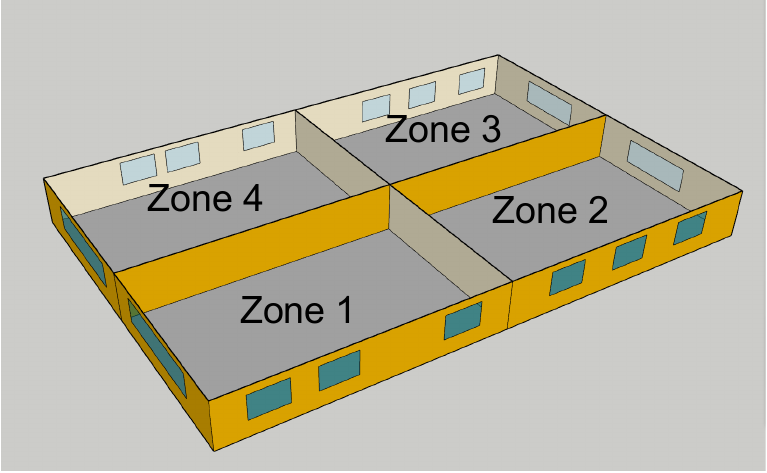}
    \caption{The building shape used in the TRNSYS for simulation.}
    \label{room}
\end{figure}

In this section, we study the performance of the proposed event-triggered encrypted control framework using the TRNSYS building simulator. To this end, we built a single-story house with four zones using TRNSYS, as shown in Fig. \ref{room}. The total floor size of the building is $360$ square meters, divided into four equal-sized zones. Each room's external facade has four windows. A heating, ventilation, and air-conditioning (HVAC) system was used to control the CO$_2$ concentration and temperature of each zone of this building. We used the Fresno City weather data from July $1$st $2016$ to July $31$st $2016$ for simulation, and the sampling time of the simulation was 5 minutes. \textcolor{black}{ We used $23.5$ degrees Celsius as the reference temperature and $800$ ppm as the reference CO$_2$ level. The occupancy of each zone in our simulation was between $1$ and $8$, and the threshold level $T_s$ for the event-triggering unit varies from $7$ to $20$.}

We implemented the event-triggered MPC using PyTorch in Python, and the CKKS encryption/decryption was implemented using the Microsoft TenSEAL library in Python. According to the official benchmark of TenSEAL, we set the CKKS parameters as polynomial degree $N=8192$, coefficient modulus sizes $q= [40,\ 26,\ 26,\ 26,\ 40]$, and the scaling factor was set to $2^{26}$. The TRNSYS-Python communication was established using TRNSYS type $3157$, which calls a Python module implemented in a .py file located in the same directory as the TRNSYS input file (the deck file) at each iteration. We implemented all simulations in Python and TRNSYS on an Intel Core i$7$-$10700$ CPU $2.9$ GHz computer.

We used the REINFORCE \cite{williams1992simple} algorithm to compute the optimal event-triggering policy. In our numerical results, the event-triggering unit was composed of two fully connected layers with a hidden unit size of $100$ to optimize the objective function of (\ref{objective function}). We ran the algorithm with enough episodes to achieve convergence.
\subsection{Benchmark}
We compare the performance of the proposed optimal event-triggered scheme with that of a threshold-based event-triggered control method with the following triggering condition 
\begin{equation}
    \label{threshold}
    \begin{aligned}
    ||\bm{e_t}||_{\infty}= ||\bm{x}_{t}-\bm{x}_{s_t^\star}||_{\infty}> \alpha,     
    \end{aligned}
\end{equation}
where $\bm{x}_{s_t^\star}$ is the last state communicated with the cloud-based controller. This event-triggering unit compares the infinity norm of the error with the predefined threshold $\alpha$, and it communicates the (encrypted) state to the cloud if the error is greater than the threshold value. Then, the cloud-based controller computes the air mass flow using the encrypted MPC policy.

\subsection{Performance Metrics}
 We use two metrics to study the impact of different event-triggering policies on the closed-loop performance: $(i)$ the percentage of violations, and $(ii)$ the maximum violation. The total violation for temperature is defined as the percentage of time instances where the temperature of at least one zone is outside the comfort temperature band $22$-$25$ Celsius. Similarly, the percentage of violations for CO$_2$ is defined as the percentage of time instances that the CO$_2$ level of at least one zone is above the concentration of \textcolor{black}{$800$} ppm. The maximum violation for temperature is defined as the value of the largest deviation of temperature from the comfort band among all zones and all time instances. Similarly, the maximum deviation for CO$_2$ is the largest deviation of CO$_2$ from the comfort level of \textcolor{black}{$800$} ppm among all zones and all time instances.

 We use the communication rate as a metric to study the impact of the optimal and threshold-based event-triggering policies on the communication frequency between the system and the cloud. The communication rate is defined as the percentage of time instances an even-triggering unit communicates with the cloud-based controller.

\begin{figure*}[ht]
\centering
\subfigure[The percentage of violations for temperature under the optimal and threshold-based event-triggering policies as a function of the communication rate.]{\label{violation_temp}\includegraphics[width=2.8in]{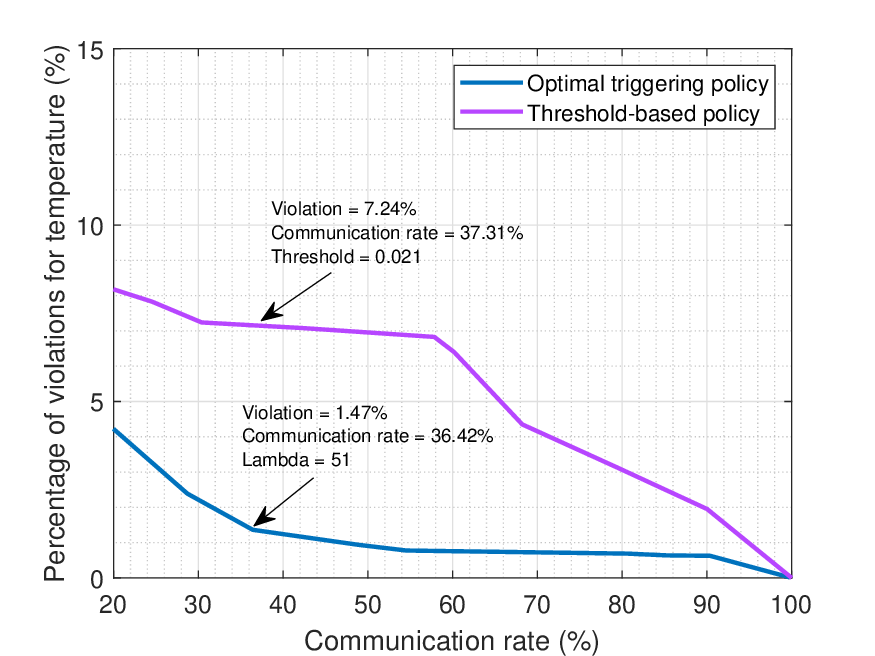}}
\quad
\subfigure[The maximum temperature violation for different communication rates under the optimal and threshold-based triggering policies.]{\label{deviation_temp}
\includegraphics[width=2.8in]{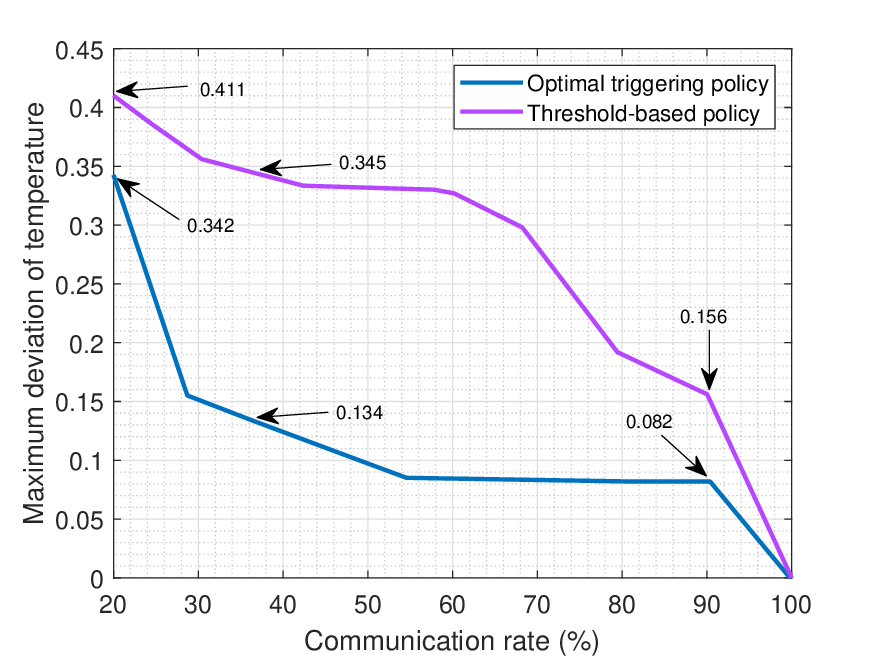}
}
\quad
\subfigure[The percentage of violations for CO$_2$ level under the optimal and threshold-based event-triggering policies as a function of the communication rate.]{\label{violation_co2}
\includegraphics[width=2.8in]{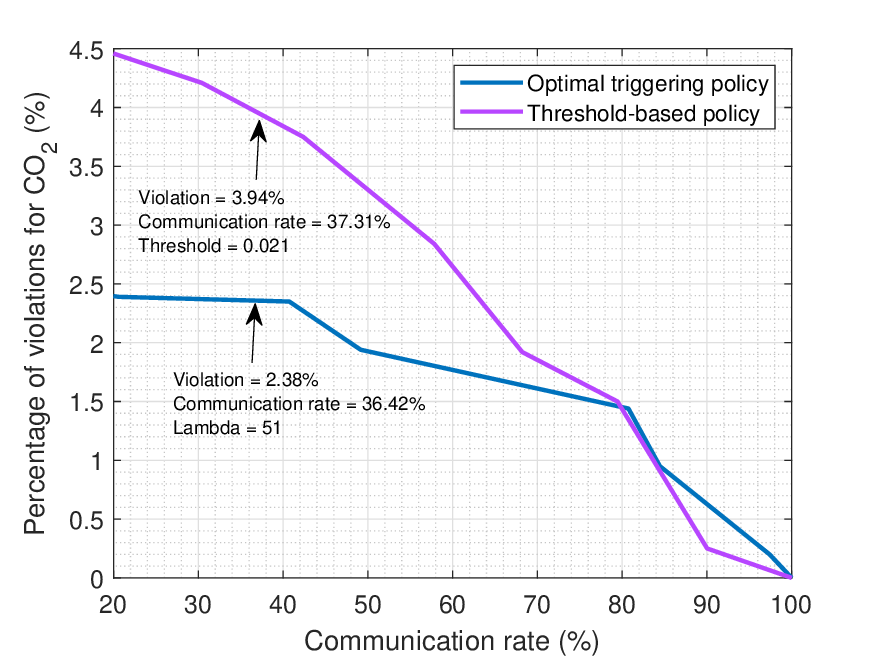}
}
\quad
\subfigure[The maximum CO$_2$ violation for different communication rates under the optimal and threshold-based triggering policies.]{\label{deviation_co2}
\includegraphics[width=2.8in]{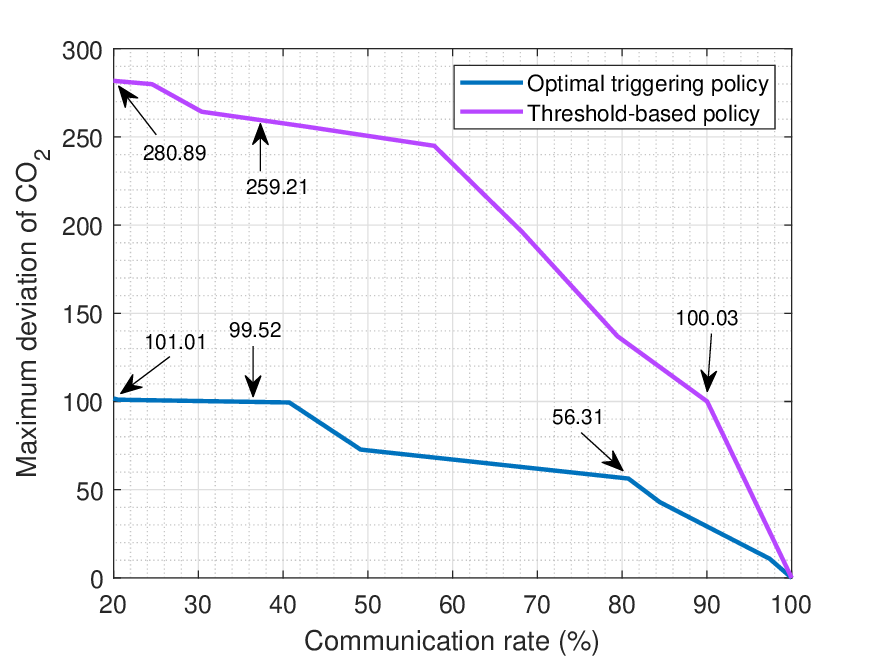}
}
\caption{The performance of the HVAC system in regulating the temperature and CO$_2$ under the optimal and threshold-based event-triggering policies.}
\end{figure*}

\subsection{Communication Rate and Control Performance Trade-off}
In this subsection, we study the trade-off between the control performance and the communication rate. Fig. \ref{violation_temp} illustrates the percentage of violations for temperature under the optimal and threshold-based event-triggering policies as a function of the communication rate. To achieve different communication rates, we varied $\lambda$ for the optimal \textcolor{black}{event-triggering} policy and $\alpha$ for the threshold-based triggering policy. For each value of $\lambda$ ($\alpha$), we computed the violations based on the performance of the HVAC system in regulating the CO$_2$ and temperature for $30$ days.

According to Fig. \ref{violation_temp}, the percentage of violation for temperature increases as the communication rate decreases under the optimal and threshold-based event-triggering policies. This is due to the fact that the controller receives the state less often as the communication rate decreases. Thus, the HVAC system uses control inputs which are calculated based on outdated information. Hence, the control inputs become less effective in regulating the temperature as the communication rate decreases. However, the percentage of violations is significantly lower under the proposed optimal event-triggering policy than the threshold-based policy. \textcolor{black}{For instance, the percentage of violations under the optimal policy is $5$ times smaller than that under the threshold-based policy when the communication rate is around $37$\%.}   

Fig. \ref{deviation_temp} shows the maximum violation of temperature under the optimal triggering and the threshold-based policy as a function of the communication rate. According to this figure, the maximum \textcolor{black}{deviation} increases as the communication rate becomes small. However, the proposed optimal triggering policy achieves a smaller deviation compared with the threshold-based policy. \textcolor{black}{For example, the maximum violation under the optimal policy is $2.5$ times smaller than that under the threshold-based policy when the communication rate is around $37$\%.}

\textcolor{black}{The percentage of violations and maximum violations for CO$_2$ are shown in Fig\textcolor{black}{.} \ref{violation_co2} and Fig. \ref{deviation_co2}, respectively.  According to these figures, the percentage of CO$_2$ violations and the maximum CO$_2$ violation increase as the communication rate decreases. However, the optimal event-triggering policy results in a smaller CO$_2$ violation than the threshold-based policy. }
\begin{figure}[H]
    \centering
    \includegraphics[width=2.8in]{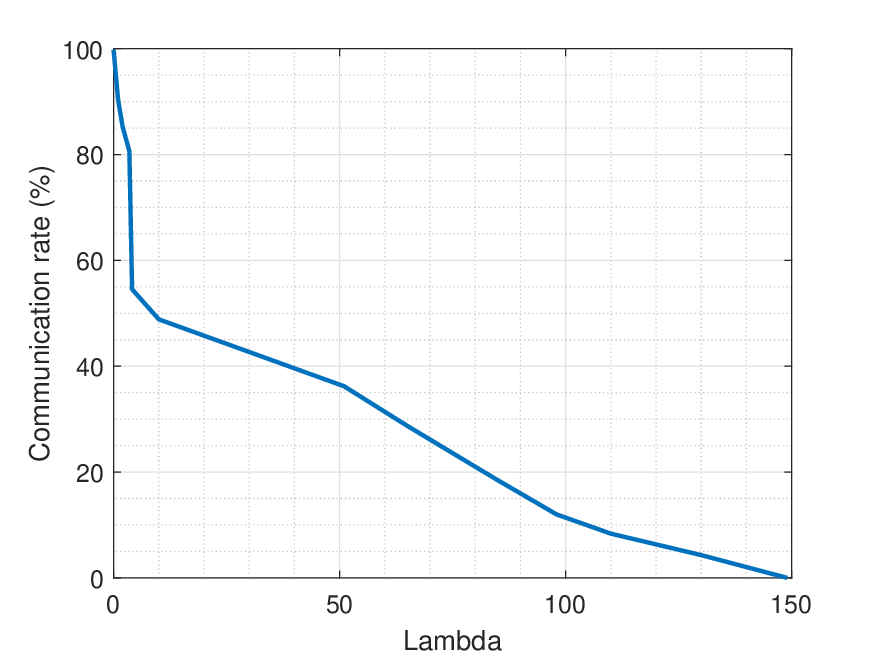}
    \caption{The communication rate of the optimal triggering policy as a function of $\lambda$.}
    \label{communication rate}
\end{figure}
Finally, Fig. \ref{communication rate} shows the communication rate of the optimal triggering policy as a function of the communication penalty ($\lambda$). According to this figure, the event-triggering unit communicates less often with the cloud as the communication penalty becomes large, which results in a larger percentage of violations and maximum violations. Based on the figures Fig. \ref{violation_temp}-\ref{deviation_co2}, the percentage of violations and the maximum violation are relatively large when the communication rate is less than $36.42\%$, which corresponds to $\lambda > 51$. We believe the $\lambda$ = $51$ leads to an acceptable balance between the control performance and communication rate.

\subsection{Impact of the MPC horizon on the Communication-Control Trade-off}
\begin{figure}[ht]
\centering
\subfigure[]{ \label{MPCvio}\includegraphics[width=2.8in]{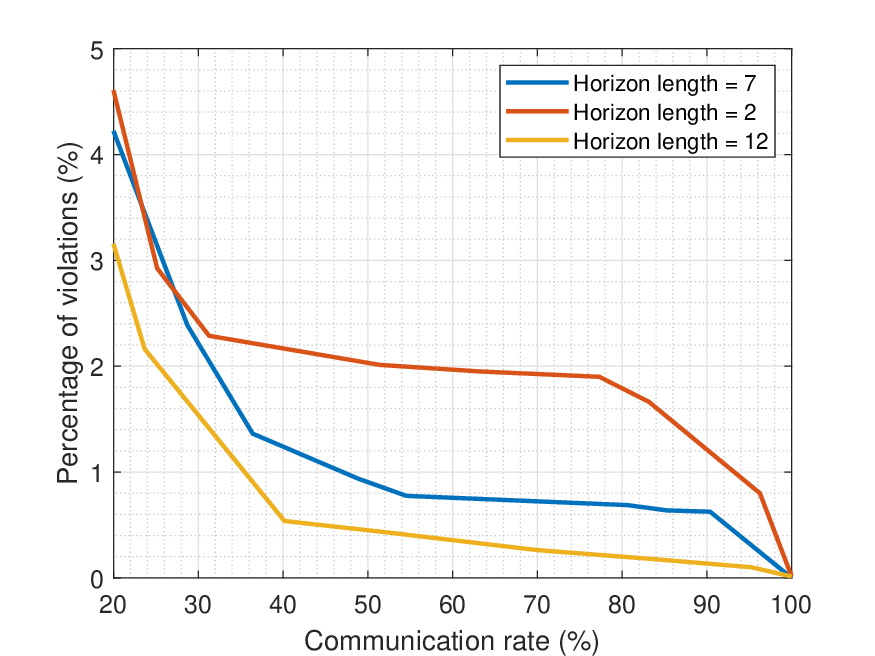}}
\subfigure[]{ \label{MPCde}
\includegraphics[width=2.8in]{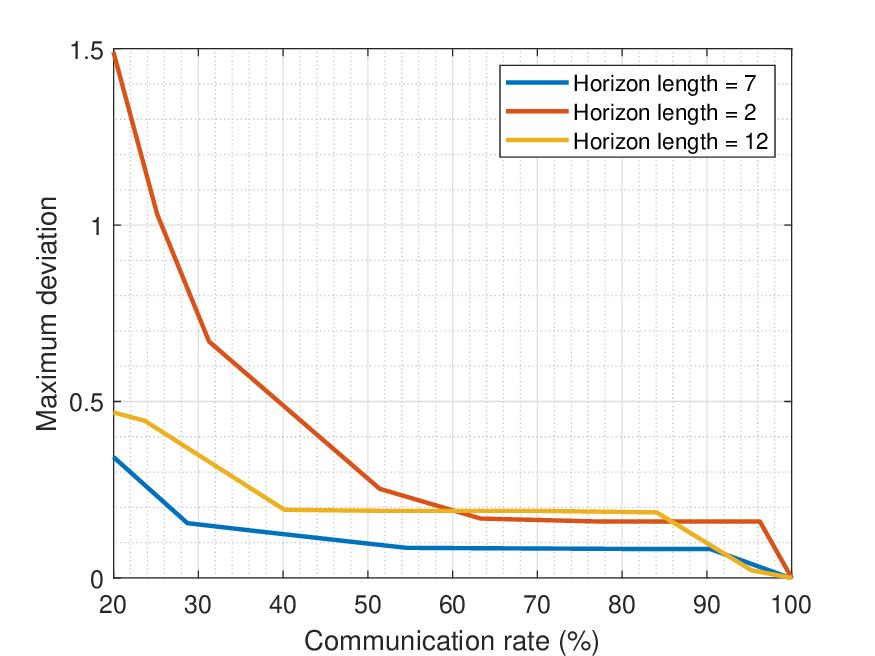}
}
\caption{The percentage of temperature violations $(a)$ and the maximum temperature violations $(b)$ under the optimal triggering policy for different horizon lengths.}
\end{figure}

In this subsection, we study the impact of the horizon length of the MPC on the communication-control trade-off under the optimal event-triggering policy. Fig. \ref{MPCvio} shows the percentage of temperature violations as a function of the communication rate for different horizon lengths under the optimal triggering policy. Based on Fig. \ref{MPCvio}, the percentage of violations for the horizon length $12$ is $0.5\%$ and $ 4.5\%$ lower than the violations when the horizon length is $7$ and $2$, respectively.

Fig. \ref{MPCde} shows the maximum temperature deviation under different horizon lengths for different communication rates. Based on this figure, the horizon length $7$ results in the smallest maximum deviation. Also, the maximum deviation for length $12$ is slightly larger than length $7$. The maximum deviation under length $2$ is close to the length $12$ when the communication rate is more than $60\%$. However, the maximum deviations under the length $12$ increases rapidly when the communication rate is below $60\%$. Based on Figs. \ref{MPCvio} and \ref{MPCde}, the performance difference between horizon lengths $7$ and $12$ is not significant. However, the MPC computation time for length $7$ is almost half the computation time for length $12$. Hence, we selected the length $7$ in our numerical results.

\subsection{Impact of the number of FGM iterations on the Communication-Control Trade-off}\label{iterationexp}
In this subsection, we study the impact of the number of iterations of FGM on the communication-control trade-off. Applying Algorithm \ref{FGM-Enc} to solve the optimization problem \eqref{MPC} requires a few rounds (iterations of Algorithm \ref{FGM-Enc}) of communication between the system and the cloud. Although the encryption burden increases with the number of iterations, in our case, an acceptable control performance can be attained by using only one iteration of the encrypted FGM. To demonstrate this observation, we numerically studied control performance under different iterations of FGM algorithm, where we chose $\lambda=51$ which corresponds to communication rate $=37\%$, and we varied the number of iterations from 1 to 5.

\begin{figure}[htb]
\centering
\includegraphics[width=2.8in]{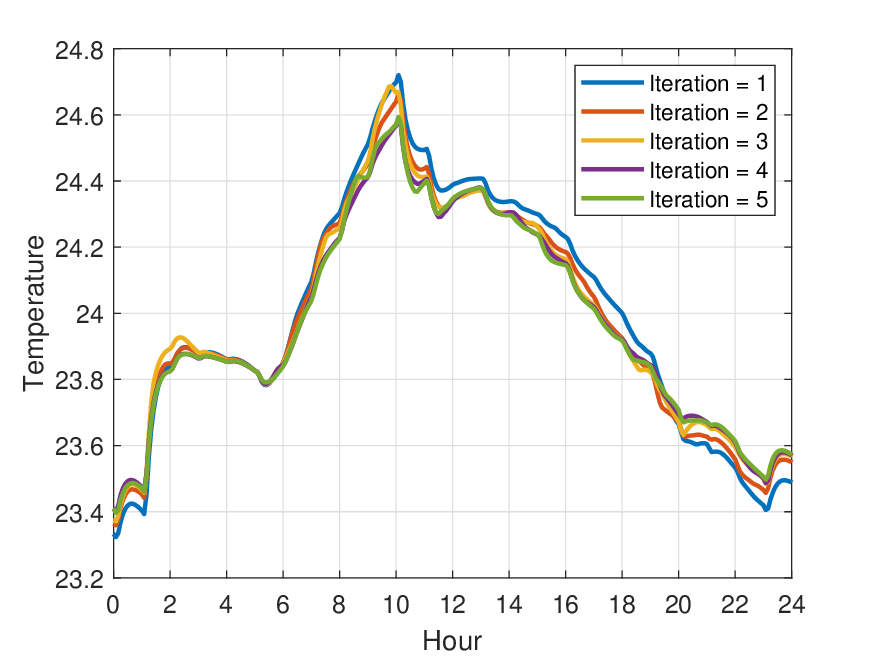}
\caption{The temperature of zone 1 of the building under different iterations of the encrypted FGM.}
\label{iteration}
\end{figure}

Fig. \ref{iteration} shows the temperature of zone 1 in 24 hours under different number of iterations of the FGM algorithm. As shown in Fig. \ref{iteration}, the number of iterations has a small impact on the performance of the HVAC system. We also studied the percentage of violations for temperature under different iteration numbers for $\lambda=51$. Here, our results indicate that the percentage of violations is 1.47\%, 1.24\%, 1.22\%, 1.15\% and 1.13\% when the number of iterations ranges from 1 to 5, respectively. According to these results, one iteration of the encrypted FGM can ensure a desirable control performance.

\subsection{Control Performance for a Fixed Communication Rate}
\begin{figure}[htb]
    \centering
    \includegraphics[width=2.8in]{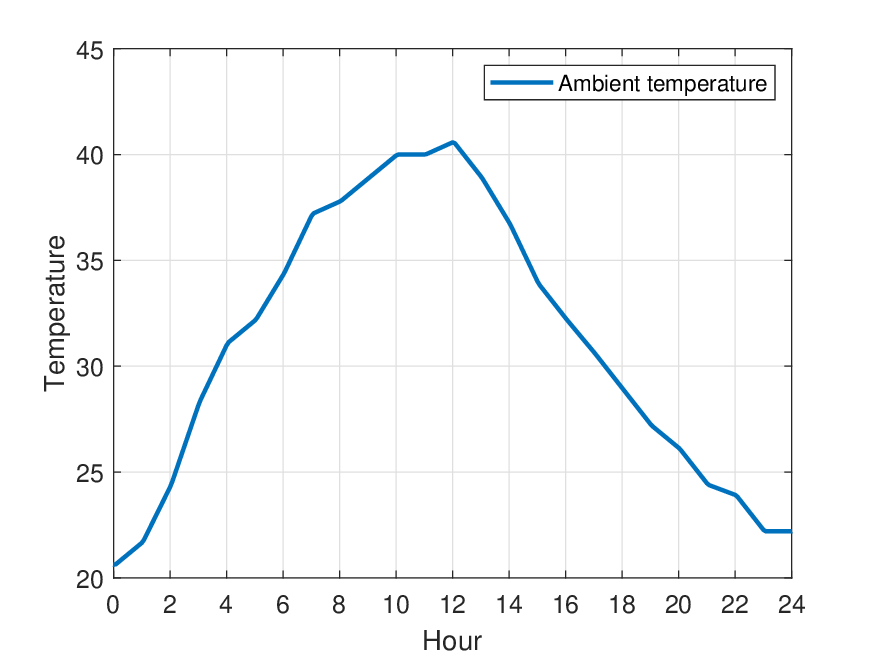}
    \caption{The ambient temperature in one day.}
    \label{ambient}
\end{figure}

\begin{figure}[htb]
    \centering
    \subfigure[]{\label{temrein}
    \includegraphics[width=2.8in]{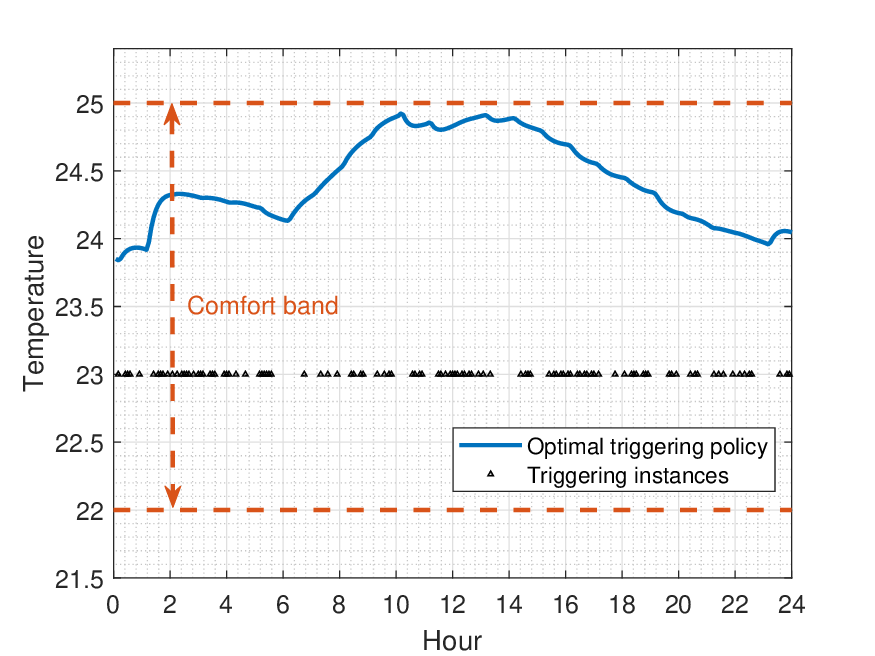}}
    \quad
    \subfigure[]{\label{temthres}
    \includegraphics[width=2.8in]{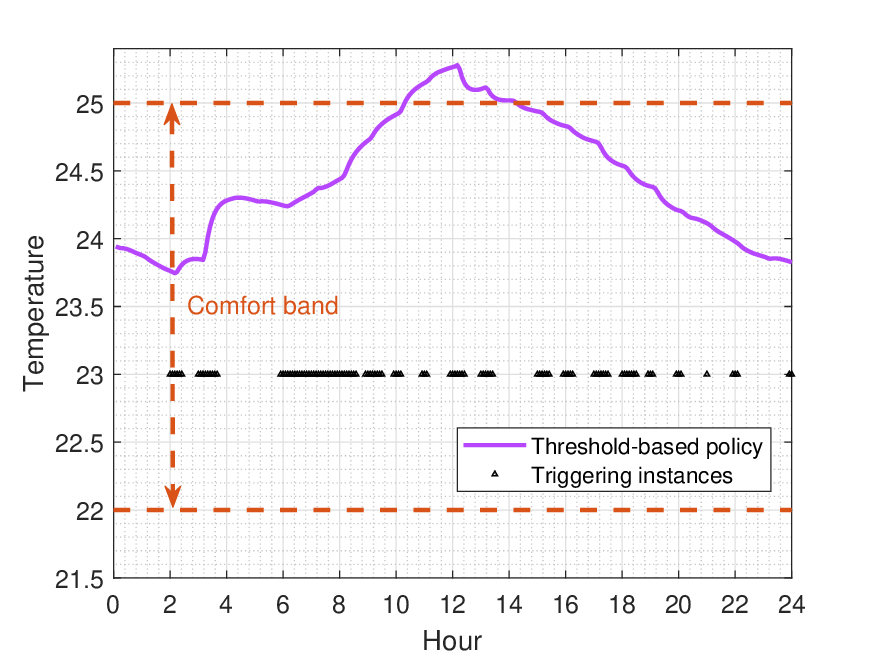}}
    \caption{The temperature of zone $1$ of the building under the optimal policy $(a)$ and threshold-based event-triggering policy $(b)$.}
    \label{daytemall}
\end{figure}

In this subsection, we compare the performance of the HVAC system in regulating the temperature and CO$_2$ (in one day) under the optimal and threshold-based policies when the communication rate is $37\%$. Fig. \ref{ambient} shows the ambient temperature during the day, which we used in our simulations. The selected day is particularly suitable for studying the performance of the event-triggering policies, since the large temperature variations act as a severe disturbance.   
\textcolor{black}{Fig. \ref{daytemall} shows} the temperature of zone $1$ of the building under the optimal and threshold-based event-triggering policies when the outdoor temperature changes according to Fig. \ref{ambient}. Based on this figure, the optimal event-triggering policy ensures that the temperature stays within the comfort band all the time despite the disturbance due to ambient temperature. However, the temperature under the threshold-based policy violates the comfort band when the ambient temperature reaches its maximum value. This observation confirms that the proposed optimal event-triggering strategy is more efficient in regulating the temperature than the threshold-based policy.
The triggering instance of the event-triggering unit under the optimal and threshold-based policies are also shown in Figs. \ref{temrein} and \ref{temthres}, respectively. Under the threshold-based policy, the event-triggering unit only communicates the cloud when the temperature changes significantly which results in concentrated triggering instances. Therefore, the system may not communicate with the cloud for a long period of time, resulting in the violation of the thermal comfort band. However, triggering instances are more evenly distributed under the optimal triggering policy. This ensures that the controller receives the state more often.

\begin{figure}
    \centering
    \includegraphics[width=2.8in]{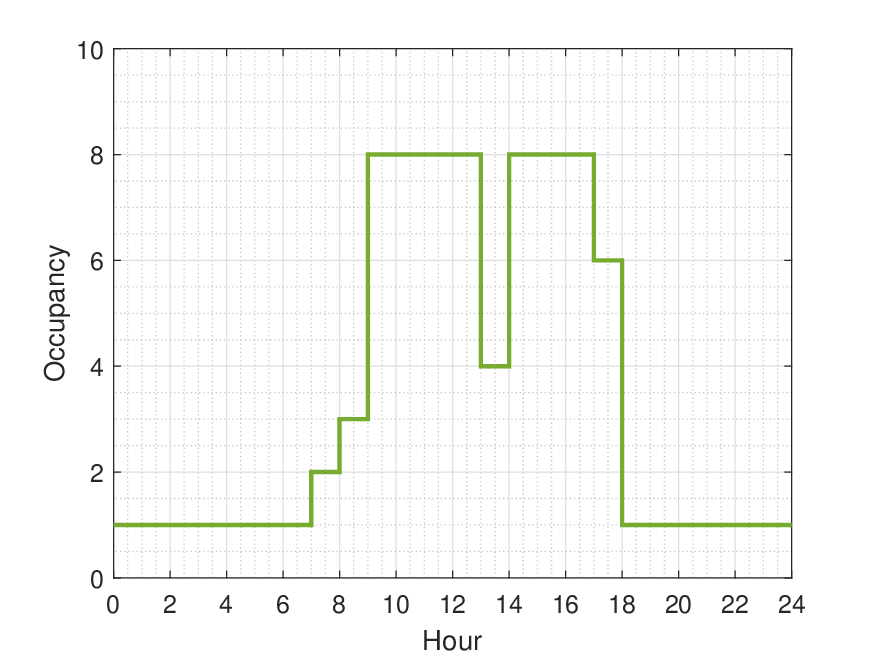}
    \caption{The occupancy schedule of Zone 1.}
    \label{roompeo}
\end{figure}

\begin{figure}
    \centering
    \includegraphics[width=2.8in]{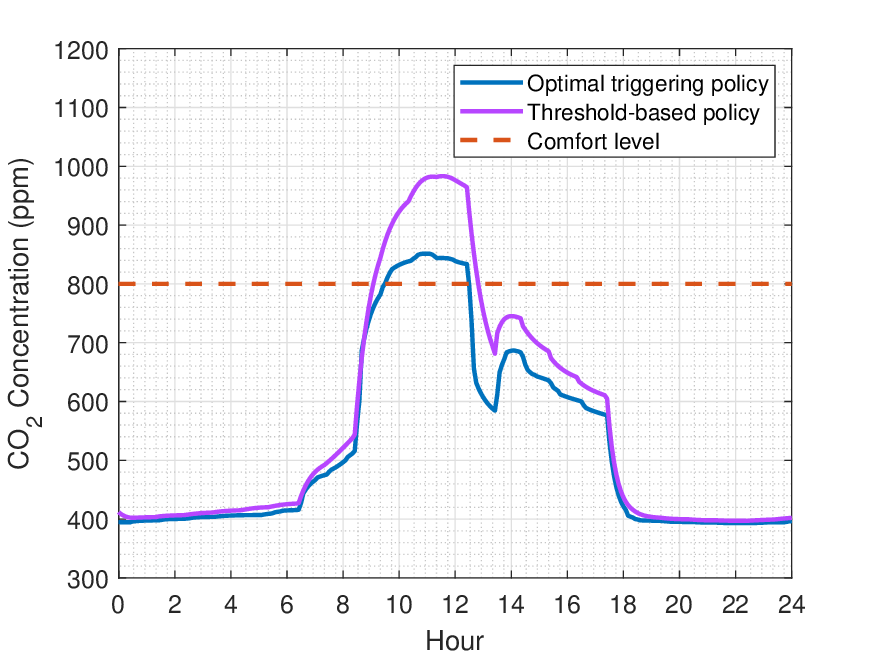}
    \caption{The CO$_2$ concentration level of Zone 1 under the optimal and threshold-based policies.}
    \label{CO2graph}
\end{figure}

We next study the behavior of indoor CO$_2$ in zone $1$ of the building under the event-triggering policies. To this end, we assumed that the occupancy of zone $1$ varies according to Fig. \ref{roompeo}, which indicates a large variation in the occupancy level. Fig. \ref{CO2graph} shows the behavior of CO$_2$ in zone $1$ under the optimal and threshold-based event-triggering policies. According to this figure, the indoor CO$_2$ increases as the occupancy becomes large. Under the threshold policy, the CO$_2$ level violates the comfort band for a longer period of time compared with the optimal triggering policy. Also, the maximum CO$_2$ violation under the threshold-based policy is larger than that under the optimal triggering policy. These observations confirm that the optimal event-triggering policy is more efficient in regulating the \textcolor{black}{indoor CO$_2$ level.}

\subsection{ Communication and Computation Performance }

\begin{figure}[h]
    \centering
    \includegraphics[width=3in]{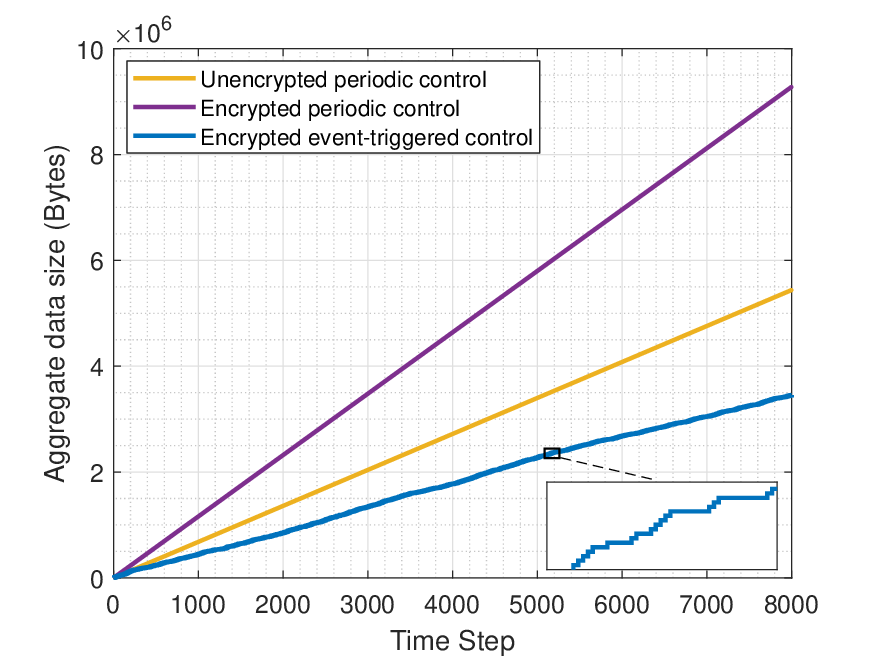}
    \caption{The size of total communicated data between the plant and cloud under the encrypted period control, unencrypted periodic control, and optimal encrypted event-triggering control schemes.}
    \label{size}
\end{figure}
In Fig. \ref{size}, we compare the total size of data, which was communicated between the plant and cloud, as a function of time in three cases: $(i)$ unencrypted periodic control, $(ii)$ encrypted periodic control, and $(iii)$ the optimal event-triggering policy with $\lambda=51$. According to this figure, the encrypted periodic control increases the size of communicated data by $170\%$ compared with the unencrypted periodic control. This is particularly a major issue when wireless sensors are used since the communication unit of a sensor accounts for the majority of power consumption of a wireless sensor \cite{yadav2016review}. Based on Fig. \ref{size}, the data size under the optimal event-triggering encrypted control is around $40\%$ of the data size under the encrypted periodic control, which indicates a significant reduction in data size and power consumption by sensors.
    \begin{figure*}[ht]
    \centering
    \subfigure[The temperature of zone 1 as a function of time under different policies.]{\label{twostrategiestem}\includegraphics[width=2.8in]{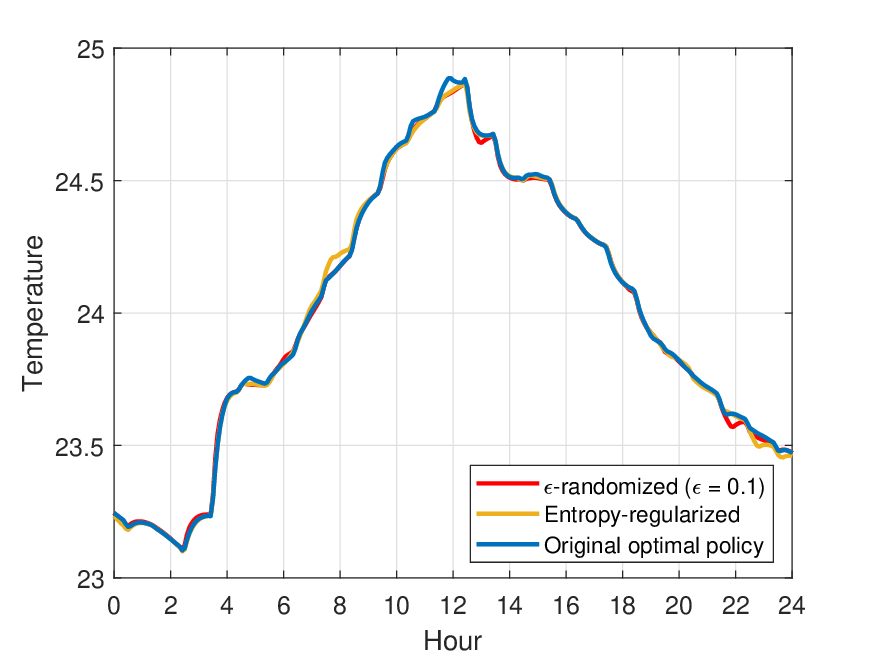}}
    \quad
    \subfigure[The CO$_2$ concentration of zone 1 as a function of time under different policies.]{\label{twostrategiestemco2}
    \includegraphics[width=2.8in]{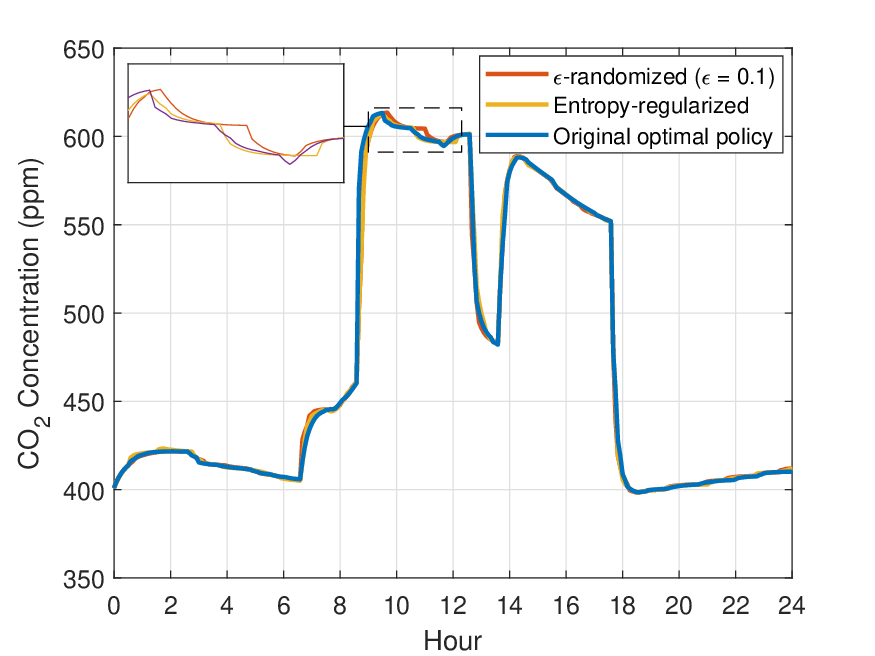}
    }
    \caption{The performance of the HVAC system in regulating the temperature and CO$_2$ under the $\epsilon$-randomized, entropy-regularized and original optimal policies.}\label{strategy1}
    \end{figure*}

    \begin{figure*}[ht]
    \centering
    \subfigure[The percentage of violations for temperature under different policies as a function of the communication rate.]{\label{twostrategiesvio}\includegraphics[width=2.8in]{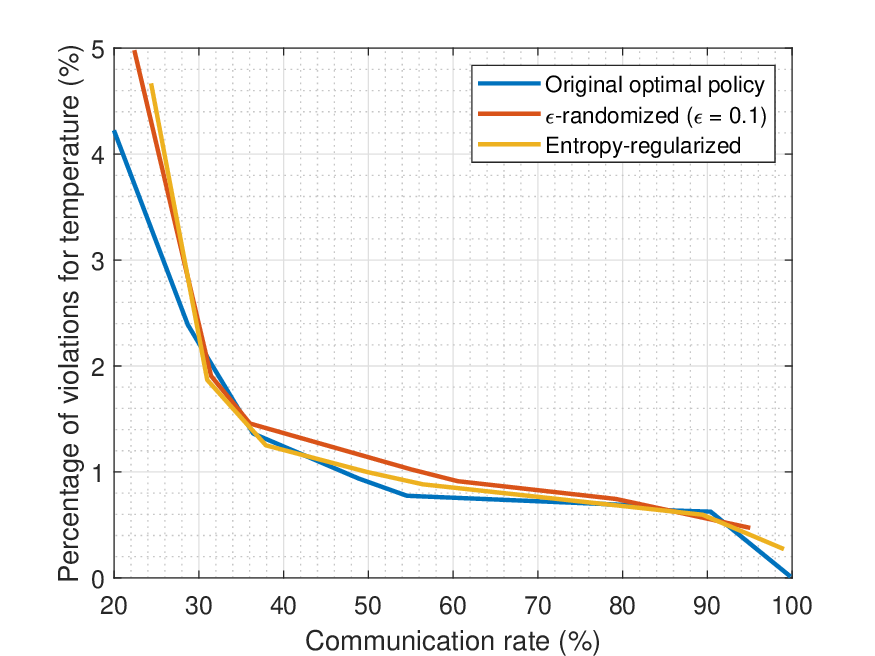}}
    \quad
    \subfigure[The maximum temperature violation for different communication rates under different policies.]{\label{twostrategiesde}
    \includegraphics[width=2.8in]{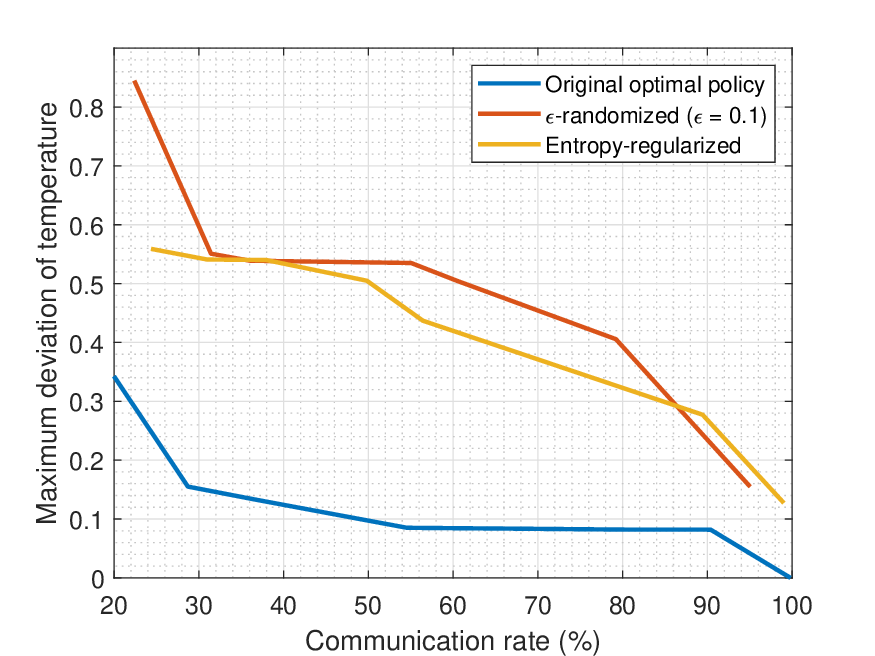}
    }
    \caption{The percentage of violations and maximum deviation of the HVAC system in regulating the temperature under the $\epsilon$-randomized, entropy-regularized and original optimal policies.}\label{strategy2}
    \end{figure*}
In our numerical simulations, it took $0.01$ seconds to solve the unencrypted MPC, whereas solving the encrypted MPC required $3.62$ seconds. This observation shows that the encrypted control requires more computational resources than the unencrypted control. This is particularly important when computing-as-a-service platforms are utilized to implement the cloud controllers, as the cost of using such platforms depends on the amount of computational resources utilized by the client. The event-triggered communication can reduce the required computational resources for the encrypted control, as the control input is only updated at the triggering instances. For instance, in our numerical results, the event-triggering unit reduced the required computational resources for solving the encrypted MPC by more than $60\%$.

\subsection{Control Performance for Randomized Triggering Policies}

    Fig. \ref{strategy1} shows the performance of the HVAC system in controlling the temperature and CO$_2$ under the $\epsilon$-randomized, entropy-regularized and original optimal policies. Based on this figure, the performance of the HVAC system under the $\epsilon$-randomized and the entropy-regularized policies is close to that under the original optimal policy. Fig. \ref{strategy2} shows the percentage of violations and maximum deviation of the temperature under the $\epsilon$-randomized, entropy-regularized and original optimal policies. According to this figure, the percentage of violations of the randomized strategies are similar, and the entropy-regularized achieves a smaller maximum deviation overall compared with the $\epsilon$-randomized policy. However, the two strategies will slightly decrease the control performance as they deviate from the optimal policy.

\section{Conclusion}
In this paper, we developed an encrypted control framework for privacy-preserving cloud-based HVAC control. In our framework, the sensor measurements of the HVAC system are encrypted using FHE scheme prior to communication with the cloud. We developed an encrypted fast gradient algorithm that allows the cloud to compute the control inputs based on encrypted sensor measurements. We also developed an optimal event-triggering policy to reduce the communication and computation costs of FHE. To reduce the leakage of private information via the triggering instances, we proposed two stochastic event-triggering policies. We finally studied the performance of the developed encrypted HVAC control framework using the TRNSYS simulator. Our future work includes the experimental study of encrypted event-based HVAC control using an HVAC testbed.

\appendices 
\section{Thermal Model of the Building}\label{App: BM}
We use an RC network to model the thermal dynamics of a building, where walls and rooms are modeled as separate nodes. The temperature of the $j$-th ($j = 1, 2, ...,n$) external wall is governed by the following equation:
\begin{equation}
    \label{wall}
    \frac{T_{o}-T_{w}^j}{R_{x}^j+\frac{R_j}{2}} +\frac{T_{r}^i-T_{w}^j}{R_{n}^j+\frac{R_j}{2}}+\alpha A_{j} Q_{r}^j=C_{j}\frac{dT_{w}^j}{dt},
\end{equation}
where $T_{w}^j$ is the temperature of wall $j$, $T_r^i$ is the temperature of room $i$ which is adjacent to wall $j$, $T_{o}$ is the outside temperature.
The parameters $R_{n}^j$ and $R_{x}^j$ are the thermal resistance for convection on the internal and external sides of the wall $j$, $R_j$ is the total thermal resistance of the wall $j$, $\alpha$ is the absorptivity coefficient, $Q_{r}^j$ is the total radiation heat that reaches wall $j$, $A_j$ is the area of wall $j$, and $C_{j}$ is the thermal capacity of wall $j$.

If wall $j$ is an internal wall, its temperature is governed by 
\begin{equation}
    \label{wall2}
    \sum_{i\in \mathfrak{W}_j}\frac{T_{r}^i-T_{w}^j}{R_{n}^j+\frac{R_j}{2}}=C_{j}\frac{dT_{w}^j}{dt},
\end{equation}
where $\mathfrak{W}_j$ is the set of rooms which are 
adjacent to wall $j$.

The temperature of the $i$-th ($i = 1, 2, ...,n$) room is governed by the following equation:
\begin{equation}
    \label{roomthermal}
    \sum_{j\in \mathfrak{R}_i}\frac{T_{w}^j-T_{r}^i}{R_{n}^j+\frac{R_j}{2}} +m^i_t C_p(T_a^i-T_r^i)+ Q_{h}^i=C_{i}\frac{dT_{r}^i}{dt},
\end{equation}
where $m^i_t$ is the air mass flow of room $i$ at time $t$, $C_p$ is the specific heat of air (the amount of heat required to raise its temperature by one degree), $T^i_a$ is the supply air temperature, $Q_{h}^i$ is the heat generation inside room $i$ (\emph{e.g.}, from human, lighting), $C_{i}$ is the thermal capacity of room $i$ and $\mathfrak{R}_i$ is the set of walls
adjacent to room $i$. The thermal model in \eqref{thermal} is obtained by combining the equations \eqref{wall}-\eqref{roomthermal}.

\section{A Brief Description of the CKKS Encryption Scheme}\label{App: CKKS}
CKKS is a fully HE technique that allows additive and multiplicative computation based on encrypted data. In the CKKS scheme, an encoder transforms the plaintext data into polynomials before encryption. After encoding, secret and public keys are used to encrypt and decrypt the message.
Let $s$ denote the secret key under the CKKS scheme, which is a polynomial. Then, the public key is generated according to 
\begin{equation}
    \label{gen}
    p=(-{a}\cdot{s}+{e},a),\nonumber
\end{equation}
where $a$ is a randomly selected number and $e$ is a noise term.

Under CKKS, the plaintext vector $l$ is encrypted as 
\begin{equation}
    \begin{aligned}\label{encrypt}
    \enc{{l}} &=({l},{0})+{p}\nonumber \\
    &=\left({c}_0,{c}_1\right)\nonumber,
    \end{aligned}
\end{equation}
where $\left({c}_0,{c}_1\right)$ denotes encrypted version of $l$. Let $c=\left({c}_0,{c}_1\right)$ and $c^\prime=\left({c}^\prime_0,{c}^\prime_1\right)$ denote the encryption of $l$ and $l^\prime$, respectively. Then, the encrypted addition of $l$ and $l^\prime$ is performed as 
\begin{align}
    \label{enadd}    \enc{{l}}\oplus\enc{l^\prime}&=c+c^\prime,\nonumber\\
    &=\left(c_0+c_0^\prime,c_1+c^\prime_1\right)\nonumber.
\end{align}   
The encrypted multiplication of $l$ and $l^\prime$ involves two operations, where we first perform the following ciphertext-ciphertext multiplication
\begin{equation}
    \label{enmult}
\enc{{c}}\otimes\enc{{c'}}=\left({c}_0\times{c_0'},{c}_0\times{c_1'}+{c_0'}\times{c_1},{c}_1\times{c_1'}\right).\nonumber
\end{equation}   
followed by a linearization step using an evaluation key \cite{fan2012somewhat}. Finally, under CKKS, the decryption is performed as 
\begin{align}
    \label{decrypt}
    \dec{{c}} &= \enc{{c}_0,{c}_1}(1,{s})^\top\nonumber\\
    &={{c}_0}+{{c}_1}\times{s}.\nonumber
\end{align}

\section{Proof of Theorem \ref{Theo: Suff-Inf}}\label{App: Suff-Inf}
We prove this result by induction. According to \eqref{objective function}, the optimal value function at time $t=H+1$ can be written as 
\begin{equation}
\label{H+1}
V_{H+1} (\mathfrak{I}_{H+1}) = (\bm{x}_{H+1}-\bm{x}_r)^\mathrm{T} \bm{Q} (\bm{x}_{H+1}-\bm{x}_r). \nonumber
\end{equation}
 Thus, $\bm{x}_{H+1}$ is the necessary information for characterizing the optimal value function at time $t=H+1$. Now, suppose that the optimal value function at time $k+1$ is a function of $\bm{x}_{k+1}$, $\bm{x}_{s_{k+1}^\star}$ and $L_{k+1}$. We show that the optimal value function at time $k$ is a function of $\bm{x}_k$, $\bm{x}_{s_k^\star}$ and $L_k$ as follows. The optimal value function at time $k$ can be written as \eqref{valuek}.
\begin{figure*}
\begin{align}\label{valuek}
V_k (\mathfrak{I}_k) = \min_{a_k }\left[(\bm{x}_k-\bm{x}_r)^\mathrm{T} \bm{Q} (\bm{x}_k-\bm{x}_r) + \bm{u}_{k}^\mathrm{T} \bm{R} \bm{u}_{k}
 + \lambda a_k + E \left\{ V_{k+1} (\bm{x}_{k+1},\bm{x}_{s_{k+1}^\star},L_{k+1})|\mathfrak{I}_k,a_k \right\} \right].
\end{align}
\hrule
\end{figure*}
Note that $(\bm{x}_k-\bm{x}_r)^\mathrm{T} \bm{Q} (\bm{x}_k-\bm{x}_r)$ is a function of $\bm{x}_k$. Also, $\bm{u}_k$ can be written as $\bm{u}_k=(1 - a_k)\bm{u}_{k|s_k^\star} + a_k\bm{u}_{k|k}$ where $\bm{u}_{k|k}$ depends on $\bm{x}_k$ as $\bm{u}_{k|k}$ is a static function of the state at time $k$ and $\bm{u}_{k|s_k^\star}$ depends on $L_k$ and the state at time $s_k^\star$, \emph{i.e.}, $\bm{x}_{s_k^\star}$. Moreover, we have
\begin{align}
&E\left\{V_{k+1}(\bm{x}_{k+1},\bm{x}_{s_{k+1}^\star},L_{k+1})|\mathfrak{I}_k,a_k\right\}=\nonumber \\ &\sum_{x,s,l} V_{k+1}(x,s,l)P\left(\bm{x}_{k+1}=x,\bm{x}_{s_{k+1}^\star}=s,L_{k+1}=l|\mathfrak{I}_k,a_k\right).\nonumber
\end{align}

According to \eqref{objective function}, $\bm{x}_{k+1}$ is a function of $\bm{x}_k$ and the control input $\bm{u}_k$ which depends on $\bm{x}_{k}$ $\bm{x}_{s_k^\star}$ and $L_{k+1}$ as we showed above. Also, $L_{k+1}$ depends on $L_k$ and $a_k$. Furthermore, $\bm{x}_{s_{k+1}^\star}$ depends on $\bm{x}_{k+1}$, $\bm{x}_{s_k^\star}$ and $a_k$. Thus, given $\bm{x}_k$, $L_k$, $\bm{x}_{s_k^\star}$, the random variables $\bm{x}_{k+1}$, $\bm{x}_{s_{k+1}^\star}$, $L_{k+1}$ are independent of other information in the set $\mathfrak{I}_k$, which implies that $E\left\{V_{k+1}(\bm{x}_{k+1},\bm{x}_{s_{k+1}^\star},L_{k+1})|\mathfrak{I}_k,a_k\right\}$ only depends on $\bm{x}_k$, $\bm{x}_{s_k^\star}$ and $L_k$. This observation indicates that the value function at time $k$ is only a function of $\left\{\bm{x}_k,\bm{x}_{s_k^\star},L_k\right\}$ which completes the proof.

\begin{figure*}
    \begin{align}
    \label{trajectory_loss}
   \mathscr{L}_\theta(\tau)= \sum_{t=1}^H [(\bm{x}_t-\bm{x}_r)^\mathrm{T} \bm{Q} (\bm{x}_t-\bm{x}_r) + \bm{u}_{t}^\mathrm{T} \bm{R} \bm{u}_{t}+ \lambda a_t-\beta J_\theta(\mathfrak{I}^{\star}_t)] + (\bm{x}_{H+1}-\bm{x}_r)^\mathrm{T} \bm{Q} (\bm{x}_{H+1}-\bm{x}_r).
    \end{align}
    \hrule
\end{figure*}

\section{Proof of Theorem \ref{gradient}}\label{App: gradient}
Note that $\nabla_\theta G(\theta)$ can be written as 
\begin{align}\label{Eq: Gradient}
\nabla_\theta G(\theta) &=\nabla_\theta E\left[ \mathscr{L}_\theta(\tau)\right]\nonumber\\
&= E\left[  \mathscr{L}_\theta(\tau) \nabla_\theta \log p_\theta(\tau)\right] + E\left[ \nabla_\theta \mathscr{L}_\theta(\tau)\right],
\end{align}
where $\mathscr{L}_\theta(\tau)$ is defined in \eqref{trajectory_loss}, $\tau = (\bm{x}_1, a_1, \dots, x_H,a_H,\bm{x}_{H+1})$ is a trajectory of the states and decisions of the event-triggering unit, and $ p_\theta(\tau) $ is the distribution of $\tau$ which is given by
\begin{equation}\label{Eq: trajectory_distribution}
    p_\theta(\tau) = \prod_{k=1}^H p(\bm{x}_{k+1}|a_k,\bm{x}_k,\bm{x}_{s_k^\star}) \pi_\theta(a_k|\mathfrak{I}_k^\star), 
\end{equation}
where $p(\bm{x}_{k+1}|a_t,\bm{x}_k,\bm{x}_{s_k^\star})$ is the conditional density of $\bm{x}_{k+1}$ given $a_t,\bm{x}_k,\bm{x}_{s_k^\star}$. 

Using \eqref{Eq: trajectory_distribution}, the first term in \eqref{Eq: Gradient} can be expanded as 
\begin{equation}\label{Eq: gradient-1}
E\left[  \mathscr{L}_\theta(\tau) \nabla_\theta \log p_\theta(\tau)\right] = E\left[ \mathscr{L}_\theta(\tau) \sum_{k=1}^H \nabla_\theta\log \pi_\theta(a_k|\mathfrak{I}_k^\star)
\right] .
\end{equation}
We next expand the second term in \eqref{Eq: Gradient} as follows. Note that, using the definition of $\mathscr{L}_\theta(\tau) $, we can expand  $\nabla_\theta\mathscr{L}_\theta(\tau) $ as 
\begin{align}
    \nabla_\theta\mathscr{L}_\theta(\tau) &= - \beta \sum_{k=1}^H \nabla_\theta J_\theta(\mathfrak{I}^{\star}_k). \nonumber
\end{align}
Using the definition of entropy, we have
\begin{align}
    \nabla_\theta J_\theta(\mathfrak{I}^{\star}_k)&=-[\nabla_\theta\pi_\theta(a_t=1|\mathfrak{I}^{\star}_t)\log \pi_\theta(a_k=1|\mathfrak{I}^{\star}_k)\nonumber\\
    &+\nabla_\theta\pi_\theta(a_k=0|\mathfrak{I}^{\star}_k)\log \pi_\theta(a_k=0|\mathfrak{I}^{\star}_k) \nonumber\\
    &+ \nabla(\pi_\theta(a_k=1|\mathfrak{I}^{\star}_k) + \pi_\theta(a_k=0|\mathfrak{I}^{\star}_k))]\nonumber\\
    &\stackrel{(a)}{=}-[\nabla_\theta\pi_\theta(a_k=1|\mathfrak{I}^{\star}_t)\log \pi_\theta(a_k=1|\mathfrak{I}^{\star}_k)\nonumber\\
    &+\nabla_\theta\pi_\theta(a_k=0|\mathfrak{I}^{\star}_k)\log \pi_\theta(a_k=0|\mathfrak{I}^{\star}_k)],\nonumber
\end{align}
where $(a)$ follows from
\begin{align}
\nabla(\pi_\theta(a_k=1|\mathfrak{I}^{\star}_k) + \pi_\theta(a_k=0|\mathfrak{I}^{\star}_k))&=\nabla 1\nonumber\\
& = 0.\nonumber
\end{align}
Thus, the second term in \eqref{Eq: Gradient} can be written as 
\begin{align}\label{Eq: gradient-2}
     E\left[ \nabla_\theta \mathscr{L}_\theta(\tau)\right]&= \beta E\left[\sum_{k=1}^H\nabla_\theta\pi_\theta(a_k=1|\mathfrak{I}^{\star}_t)\log \pi_\theta(a_k=1|\mathfrak{I}^{\star}_k)\right]\nonumber\\
 &+\beta E\left[\sum_{k=1}^H\nabla_\theta\pi_\theta(a_k=0|\mathfrak{I}^{\star}_k)\log \pi_\theta(a_k=0|\mathfrak{I}^{\star}_k) \right].
\end{align}
Thus, $\nabla_\theta G(\theta)$ can be obtained by combining \eqref{Eq: gradient-1} and \eqref{Eq: gradient-2}.


\bibliographystyle{IEEEtran}

\bibliography{references.bib}



\vspace{11pt}

\vspace{11pt}

\vfill

\end{document}